%% file: quad_phest_biases.tex
\documentclass[prd,amsmath,amssymb,floatfix,nofootinbib,superscriptaddress,onecolumn]{revtex4}

\usepackage{upgreek}
\usepackage{amsmath}
\usepackage{amssymb}
\usepackage{aas_macros}
\usepackage{latexsym}
\usepackage{bm}
\usepackage{stmaryrd}
\usepackage{color}
\usepackage{graphicx}
\usepackage{subfigure}
\usepackage{simplewick}
\usepackage{natbib}
\usepackage[hypertex]{hyperref}
\usepackage{ifthen}
\def\eprinttmp@#1arXiv:#2 [#3]#4@{
\ifthenelse{\equal{#3}{x}}{\href{http://arxiv.org/abs/#1}{#1}}{\href{http://arxiv.org/abs/#2}{arXiv:#2} [#3]}}

\providecommand{\eprint}[1]{\eprinttmp@#1arXiv: [x]@}
\newcommand{\adsurl}[1]{\href{#1}{ADS}}

\input{macros.tex}

\newcommand{\vn}{\mathbf{n}}

\begin{document}
\title{CMB temperature lensing power reconstruction}
\date{Draft of August 24, 2010}
\author{Duncan Hanson}
\affiliation{Institute of Astronomy and Kavli Institute for Cosmology Cambridge, Madingley Road, Cambridge CB3 OHA, UK}
\author{Anthony Challinor}
\email{adc1000@ast.cam.ac.uk}
\affiliation{Institute of Astronomy and Kavli Institute for Cosmology Cambridge, Madingley Road, Cambridge CB3 OHA, UK}
\affiliation{DAMTP, Centre for Mathematical Sciences, Wilberforce Road, Cambridge CB3 OWA, UK}
\author{George Efstathiou}
\affiliation{Institute of Astronomy and Kavli Institute for Cosmology Cambridge, Madingley Road, Cambridge CB3 OHA, UK}
\author{ Pawel Bielewicz}
\affiliation{Institut d'Astrophysique de Paris, 98bis, Boulevard Arago, 75014 Paris, France}
\affiliation{Centre d'Etude Spatiale des Rayonnements, 9 av du Colonel Roche, BP 44346, 31028 Toulouse Cedex 4, France}

\begin{abstract}
We study reconstruction of the lensing potential power spectrum from CMB temperature data, with an eye to the \Planck experiment. We work with the optimal quadratic estimator of Okamoto and Hu, which we characterize thoroughly in application to reconstruction of the lensing power spectrum. We find that at multipoles $L\!<\!250$ our current understanding of this estimator is biased at the $15\%$ level by beyond-gradient terms in the Taylor expansion of lensing effects. We present the full lensed trispectrum to fourth order in the lensing potential to explain this effect. We show that the low-$L$ bias, as well as a previously known bias at high-$L$, is relevant to the determination of cosmology and must be corrected for in order to avoid significant parameter errors. We also investigate the covariance of the reconstructed power, finding broad correlations of $\approx 0.1\%$. Finally, we discuss several small improvements which may be made to the optimal estimator to mitigate these problems.
\end{abstract}


\maketitle


\section{Introduction}
The cosmic microwave background (CMB) provides us with a picture of the Universe at the time when neutral atoms first formed, some $400,000$ years after the big bang. The $\order(10^{-5})$ anisotropies in its temperature characterize the density perturbations at that time, and from them we have been able to learn a good deal about the properties of our Universe (see Ref.~\cite{2010arXiv1001.4538K} for the most up-to-date constraints). The CMB is a powerful tool because it is largely free of astrophysical complications. Its features may be derived from initial conditions using linear physics, and to a good approximation it may be treated as a Gaussian random field statistically characterized by a single power spectrum.

At a high level of sensitivity and resolution, however, one expects to find small deviations from Gaussianity. These are expected at some level both from the high-energy physics which generated the primordial perturbation thus seeding the CMB anisotropies, and from secondary astrophysical effects after the CMB was released at recombination (see Ref.~\cite{2004PhR...402..103B} for a review). One of the most important astrophysical non-Gaussianities is that introduced by gravitational lensing of the CMB (see Ref.~\cite{LewChaLens} for a review.) We view the distant Universe through the intervening matter of our Hubble volume, the gravitational potential of which distorts the paths of the CMB photons. In essence, we view the CMB through bubbled glass. This blurs the features of the underlying anisotropies~\cite{1996ApJ...463....1S}, and introduces non-Gaussianity into the observed sky~\cite{2000PhRvD..62f3510Z}.
In some sense, lensing of the CMB is a nuisance as it provides a potential contaminant for measurements of intrinsic non-Gaussianity in the early Universe~\cite{1999PhRvD..59j3002G,2009PhRvD..80h3004H}, with which we hope to be able to discriminate between fundamental theories. From another perspective, however, CMB lensing is a tool which provides us with a new perspective on the contents of our Universe~\cite{2002PhRvD..65b3003H,2008arXiv0811.3916S}. By studying the warping effect which large scale structure has on the paths of CMB photons, we may make integrated measurements of the matter power spectrum at intermediate redshifts. This provides us with additional powers to refine parameter measurements, and to break degeneracies in measurements of the primary CMB~\cite{1999MNRAS.302..735S,2006PhRvD..73b3517S,2006PhRvD..74l3002S}. With \Planck lensing data, for example, we expect to be able to constrain the sum of neutrino masses to sub-eV levels \cite{LesPerPasPia}.
 
The effect of lensing is predominantly a simple remapping of the temperature distribution of the underlying, unlensed CMB by the lensing deflection. To leading order, the deflection is the gradient of a lensing potential $\phi$ which is an integrated measure of the intervening gravitational potential along the unlensed line of sight (e.g.~\cite{LewChaLens}). We may write the remapping on the sphere as
\begin{equation}
\ThetaL(\hat{\vn}) = \ThetaU[\hat{\vn}+\nabla\phi(\hat{\vn})] ,
\label{eqn:lremap}
\end{equation}
where $\ThetaU(\hat{\vn})$ is the unlensed CMB temperature in the direction $\hat{\vn}$ and $\ThetaL(\hat{\vn})$ is the observed, lensed temperature.\footnote{Equation~(\ref{eqn:lremap}) is rather symbolic; the remapping is by a distance
$|\nabla \phi|$ along the geodesic tangent to $\nabla \phi$ through $\hat{\vn}$~\cite{2002PhRvD..66l7301C}.}
On the scales which we shall be considering, $\phi(\hat{\vn})$ is well approximated as a Gaussian random field, and can be characterized solely by its power spectrum $\cppp{L}$ in harmonic space. This power spectrum is calculable for a given cosmology (see~\cite{LewChaLens} for details) and can be compared to measurements in order to constrain parameters.

We can gain insight into the effects of lensing by considering the Taylor expansion of Eq.~\eqref{eqn:lremap}:
\begin{equation}
\ThetaL(\hat{\vn}) = \ThetaU(\hat{\vn})+\nabla_i\phi(\hat{\vn})\nabla^i\ThetaU(\hat{\vn}) + \cdots,
\label{eqn:lexp}
\end{equation}
from which we can see that the effect of lensing is to introduce dependence on the higher derivatives of the unlensed CMB temperature into the CMB itself. These additional terms can couple to give non-Gaussian effects on the sky. 
To the extent that the lensing deflections are small, the first order term in Eq.~\eqref{eqn:lexp} is sufficient to describe the lensing (but see ~\cite{ChaLewCorr} for a discussion of circumstances where this approximation fails).
Therefore, one expects that an estimator of the form 
\begin{equation}
\widehat{\nabla \phi}(\hat{\vn}) \propto \ThetaL(\hat{\vn}) \nabla\ThetaU(\hat{\vn})
\label{eqn:deflectionest}
\end{equation}
will give a reasonable reconstruction of the deflection field. Taking the expectation value over CMB realizations we can see that if properly normalized this estimator recovers the deflection field in the mean (at first order in the lensing expansion). The reconstruction is clearly quite noisy, however, as for any particular realization there will be some spurious $\ThetaU\nabla\ThetaU$ correlation. This term is zero in the mean, but contributes significant variance to the estimator. A further issue is what field to use as a proxy for the unlensed CMB.

We can achieve a somewhat better overall reconstruction than that in Eq.~\eqref{eqn:deflectionest} by taking the correlation between filtered fields. For an
isotropic survey covering the full sky with uniform noise, the minimum-variance lensing reconstruction has been worked out by Okamoto \& Hu \cite{OkHu}. We write their estimator here as
\begin{equation}
\hat{\phi}_{LM} = A_L \intsky{ Y_{LM}^{*}(\hat{\vn}) \nabla^i \left[V(\hat{\vn}) \nabla_{i} U(\hat{\vn}) \right] },
\label{eqn:hurealspace}
\end{equation}
where $A_L$ is a multipole-dependent normalization, and $V(\hat{\vn})$ and $U(\hat{\vn})$ are the filtered maps, given as
\begin{eqnarray}
V(\hat{\vn}) &=& \sum_{LM} \frac{1}{\clexptp{\el}} \Theta_{\ell m} Y_{\ell m}(\hat{\vn})\\
U(\hat{\vn}) &=& \sum_{LM} \frac{\cttup{\el}}{\clexptp{\el}} \Theta_{\ell m} Y_{\ell m}(\hat{\vn}).
\label{eqn:filteredmaps}
\end{eqnarray}
Here, $\Theta_{\el m}$ are the multipoles of the measured sky temperature (after correcting for the instrumental beam) and $\clexptp{\el}$ is the ensemble-averaged power spectrum for the measured temperature, nominally $\cttlp{\el} + N^{TT}_{\el}$, where $\cttlp{\el}$ is the
(lensed) CMB power spectrum and $N^{TT}_{\el}$ is the power spectrum of the experimental noise. In this work as a notational aid we will tend to use $L$ to denote a multipole of the lensing potential, and $\el$ to denote a multipole of temperature. The power spectrum $\cttup{\el}$ in the filter for the $U(\hat{\vn})$ field is of the unlensed temperature. This estimator has a nice interpretation. It is a correlator of the inverse-variance-weighted temperature map $V(\hat{\vn})$ with the gradient of the Wiener reconstruction $U(\hat{\vn})$ of the unlensed CMB, which agrees with the intuition above. The inverse-variance weighting reduces the magnitude of the Gaussian noise, i.e. chance $\ThetaU\nabla\ThetaU$ correlations, which contributes variance to the estimator. The divergence operation in the integrand of Eq.~(\ref{eqn:hurealspace}) serves to extract the gradient part of the estimated deflection field.

The approach of Eq.~\eqref{eqn:hurealspace}, has already been used successfully to find evidence for lensing effects in the \WMAP data by cross-correlation with external data sets \cite{SmiZahDor, HirHoSel}. The next step in the study of CMB lensing will be an entirely internal detection. It is expected that this will be possible in the near future with data from the \Planck satellite. \Planck noise levels should enable us to measure the lensing power spectrum with $S/N>1$ over the multipole range $\el=$10--300. 
 This will prove cosmologically interesting. For example, massive neutrinos should have percent-level effects on the lensing power spectrum if the sum of their mass eigenstates is $\sim\!0.1$eV \cite{KapKnoSon}. This is close to the lower limit on the summed neutrino masses inferred from oscillation data, and so \Planck will at the very least be able to place useful upper limits \cite{LesPerPasPia}.

For the purpose of obtaining internal parameter constraints we require not $\phi$, but an estimate of its power spectrum $C_L^{\phi\phi}$. A straightforward approach to obtaining this is to form the estimator
\begin{equation}
\cppest{L} = \frac{1}{2L+1} \sum_{M=-L}^{L} \hat{\phi}_{L M} \hat{\phi}_{LM}^{*}.
\label{eqn:cppestdef}
\end{equation}
Apart from the well-known bias in this estimate associated with the statistical noise in the reconstruction of $\phi$, there are several intrinsic complications which are usually neglected. The first is that although Eq.~\eqref{eqn:hurealspace} recovers the lensing potential $\phi_{LM}$ when averaged over CMB realizations, the actual reconstruction for a single realization is sensitive to the lensing potential over a variety of multipoles.
In the power spectrum reconstruction of Eq.~\eqref{eqn:cppestdef} these modes may couple to give additional biases. The magnitude of these terms was first pointed out on the flat sky in~\cite{KesCooKam}. 

A second and more general point of concern is that Eq.~\eqref{eqn:hurealspace} only reconstructs $\phi_{LM}$ to first order. The lensing expansion of Eq.~\eqref{eqn:lexp} does not generally converge as fast as one would expect given the small size of the lensing deflections, and modern observations already test its accuracy. An accurate calculation of the lensed power spectrum $\cttlp{L}$, for example, was given in \cite{ChaLewCorr}. There, it was found that working only to leading order in $C_L^{\phi\phi}$ [i.e.\ retaining the first two non-trivial terms in the
Taylor expansion in Eq.~\eqref{eqn:lexp} which is necessary for a consistent
calculation of the power spectrum to $\mathcal{O}(C_L^{\phi\phi})$] is only accurate at the 5--10\% level relative to the true lensing corrections for multipoles $\el=$1000--2000. It is quite reasonable to find a partial breakdown of the low-order Taylor expansion in this regime --- at these multipoles the typical angular scale of the CMB fluctuations is 5--10~arcmin, whereas the magnitude of the deflection $\left|\nabla\phi\right|$ is typically 3~arcmin. Things are not quite as
bad as these arguments would suggest, since it is only the \emph{relative} displacement of points that matters for the statistics of the lensed CMB, but it is still the case that there is general need for care when working with the lensing expansion of Eq.~\eqref{eqn:lexp}. Indeed, some numerical work has already been carried out~\cite{AmbValWhi} which shows the effects of higher-order terms from Eq.~\eqref{eqn:lexp} in flat-sky power spectrum reconstruction, although they only found serious biases introduced at high multipoles where astrophysical non-Gaussianities also begin to gain importance.

Finally, related to both of the first two issues, there exist intrinsic correlations between the multipoles of the reconstructed power spectrum, even on the full sky. These are to be expected since the $\hat{\phi}$ reconstruction is non-Gaussian, being quadratic in the lensed CMB temperature (which is itself mildly
non-Gaussian). The magnitude of these correlations was investigated in the flat-sky limit in~\cite{KesCooKam}. They found negligible correlations of $\order(<10^{-4})$ for all multipole pairs inside $L<2000$. Their calculation treated the lensed CMB as Gaussian, however, and numerical verification that the non-Gaussianities introduced by lensing (which couple much more readily) are negligible would be useful.

In this paper, we use numerical simulations backed up by analytical work to check that the current understanding of the reconstruction $\cppest{L}$ is adequate. We do this for the simplest possible case: on the full sky with homogeneous noise and no secondary effects other than lensing. Our simulations are discussed in Sec.~\ref{section.simulations}. In Sec.~\ref{section.psbiases} we discuss power spectrum biases. Our principle new result is that higher-order terms from the expansion of lensing effects significantly suppress power at low-$L$ in the reconstruction of $C_L^{\hat{\phi}\hat{\phi}}$. We provide simple analytic
approximations for the bias which capture the majority of the effect seen
in our simulations.
In Sec.~\ref{section.pscovariance} we give our results on the covariance of the reconstructed power spectrum, and derive further simple formulae which accurately describe the covariances on the full sky. Finally, we conclude with a discussion of the cosmological relevance of all of these complications, and discuss improvements to the usual lensing estimator which can largely remove them.
Appendices provide technical details on the harmonic-space form of the lensing
Taylor expansion and a flat-sky calculation of the main results
in Sec.~\ref{section.psbiases}.

\section{Simulations}
\label{section.simulations}

Our simulation procedure is as follows. We start with Gaussian realizations of $\phi$ and $\tilde{\Theta}$ from spectra for a $\Lambda$CDM cosmology with $h\!=\!0.73$, $\Omega_{b}h^2\!=\!0.022$, $\Omega_{m}h^2\!=\!0.127$, $n_s=\!0.95$, $\sigma_{8}\!=\!0.743$ and three species of massless neutrinos,
up to $\el_{\rm inp}=3000$. Our lensing power spectrum $\cppp{L}$, for example, is shown in Fig.~\ref{fig.naivespectra}. We neglect the $C_L^{\ThetaU\phi}$ correlation induced on large scales by the integrated Sachs-Wolfe effect. The low-$\ell$ temperature multipoles play a negligible part in the lensing reconstruction and this choice has no effect on our results.  We proceed to lens our $\ThetaU(\hat{\vn})$ realization using the LensPix framework \cite{LewLpix}. This takes the gradient of $\phi$ using a fast spin-1 transform and remaps the temperature with the resulting deflection field, using cubic interpolation to approximate Eq.~\eqref{eqn:lremap} accurately. We obtain a lensed \healpix map at resolution $N_{\text{side}}=2048$, which we convolve with a Gaussian beam of full-width at half-maximum $\fwhm\!=\!7$~arcmin. We then add uncorrelated Gaussian pixel noise with standard deviation $\sigma_{N}\!=\!27\, \mu$K-arcmin. Finally, we deconvolve the experimental beam and harmonically transform our map back into multipole space, dropping all multipoles past $\el_{\rm max}=2750$. We drop higher multipoles because the lensed power spectrum $\cttlp{L}$ results from a convolution in multipole space of width $\Delta L \lesssim 250$, and so the lensed multipoles simulated with an $\el_{\rm inp}=3000$ cutoff are not cosmically accurate above this $\el_{\rm max}$. We have verified that for the noise level which we use the contribution of information in these high multipoles to the lensing reconstruction is sufficiently small that the location of our cutoff has negligible effect on our results. The power spectra of our lensed maps agree well with those expected, namely \cite{KnoxDet}
\begin{equation}
\clexptp{\el} = \cttlp{\el} + \left(\frac{\sigma_{N}}{T_{\text{CMB}}}\right)^2 e^{\el(\el+1)\fwhm^2/8{{\rm ln}2}},
\label{eqn:cttlexpt}
\end{equation}
where $\cttlp{L}$ is the lensed power spectrum theoretically predicted by
\camb~\cite{2000ApJ...538..473L}, using the prescription of \cite{ChaLewCorr}.

For lensing reconstruction, we perform Eq.~\eqref{eqn:hurealspace}. The gradient operations are again made with spin-1 transforms. The filtering operations are performed in harmonic space, with $\cttup{\el}$ our input unlensed power spectrum and $\clexptp{\el}$ as given above.

\section{Power spectrum biases}
\label{section.psbiases}
To determine the expectation value of $\cppest{L}$ it is most natural to work in harmonic space. Here the lensing expansion becomes somewhat more complicated, but the underlying idea is the same as in real space. Schematically, we have
\begin{equation}
\telm = {\ThetaU}_{\el \m} + \delta\telm + \delta^2\telm + \delta^3\telm +
\cdots , 
\label{eqn:lexphar}
\end{equation}
where the power of $\delta$ denotes the order in $\phi$. Simplified expressions for the expansion terms are given in Appendix~\ref{app.expansionterms}. In harmonic space, the Okamoto \& Hu \cite{OkHu} estimator becomes
\begin{equation}
\hat{\phi}_{LM}=A_L\stwols(-1)^M
\ThreeJ{\el_1}{\el_2}{L}{m_1}{m_2}{-M}g_{\el_{1}\el_{2}}(L)
\ThetaL_{\lbar{1}} \ThetaL_{\lbar{2}},
\label{eqn:huharspace}
\end{equation}
where the symbol in braces is the $3j$ symbol and from here onward we use the compact notation that
$\lbar{i} = \{\el_{i},\:m_{i}\}$.
The weight function $\gwt{1}{2}(L)$ gives the filtered maps and performs the gradient operation:
\begin{equation}
\gwt{1}{2}(L) = \frac{\fwt{1}{2}}{2\cttlp{\el_{1}, {\rm expt}}\cttlp{\el_{2}, {\rm expt}}} = \frac{\cttu{2}\bfftL{1}{2} + \cttu{1}\bfftL{2}{1}}{2\cttlp{\el_{1}, {\rm expt}}\cttlp{\el_{2}, {\rm expt}}},
\label{eq:gonetwo}
\end{equation}
where the function $\bfftL{1}{2}$ describes the rotationally-invariant part of the coupling between the three multipoles $(\el_1, \el_2, L)$ and is given by
\begin{equation}
\bfftL{1}{2} = \left[ \llponemul{L}^2 - \llponemul{\el_{1}}^2 + \llponemul{\el_2}^2 \right] \twolpomul{\el_1 L \el_2} (16\pi)^{-1/2} \ThreeJ{\el_1}{L}{\el_2}{0}{0}{0} \, .
\end{equation}
Note that $\bfftL{1}{2}$ is symmetric in its last two indices. For
later convenience, we have introduced the compact notation
\begin{eqnarray}
\llponemul{a \cdots n} &\equiv& \sqrt{a(a+1) \cdots n(n+1)}\\
\twolpomul{a \cdots n} &\equiv& \sqrt{(2a+1) \cdots (2n+1)}.
\end{eqnarray} 
In terms of these, the estimator normalization is
\begin{equation}
A_L^{-1} = \frac{1}{\twolpomul{L}^2} \sum_{\el_1 \el_2}
f_{\el_1 L \el_2}g_{\el_1\el_2}(L) ,
\label{eq:normalization}
\end{equation}
which ensures that the average of $\hat{\phi}_{LM}$ over the CMB fluctuations
\emph{for fixed lenses} is the true $\phi_{LM}$.

Averaged over realizations of the CMB and lenses, we can now see that our reconstructed power spectrum has an expectation value given by
\begin{eqnarray}
\langle \cppest{L} \rangle &=& \frac{A_L^2}{\twolpomul{L}^2} \sfourls
\sum_{M}(-1)^M
\ThreeJ{l_1}{l_2}{L}
   {m_1}{m_2}{-M}
\ThreeJ{l_3}{l_4}{L}
   {m_3}{m_4}{M}
g_{\el_{1}\el_{2}}(L)g_{\ell_{3}\el_{4}}(L)\fourpt.
\label{eqn:cppestmean}
\end{eqnarray}
Thus, we need to determine the 4-point function of the lensed CMB. The dominant term is from the disconnected part, which is derived under the assumption that the lensed CMB is a Gaussian random field on the sphere: 
\begin{equation}
N^{(0)}_L = \frac{A_L^2}{\twolpomul{L}^2} 
\sum_{\el_1\,\el_2}
[\gwt{1}{2}(L)]^2 [2 \clexptp{\el_1} \clexptp{\el_2}] = A_L , 
\label{eqn:gaussianbias}
\end{equation}
where the last equality holds only when the weights are chosen (optimally) as
in Eq.~\eqref{eq:gonetwo}.
Here we have taken up the notation that $N^{(p)}_L$ gives the power spectrum bias with order $p$ dependence on the lensing power spectrum $\cppp{L}$, exempting terms where it enters as a component of the lensed CMB power spectrum.

\begin{figure}
\begin{center}
\includegraphics[width=5in]{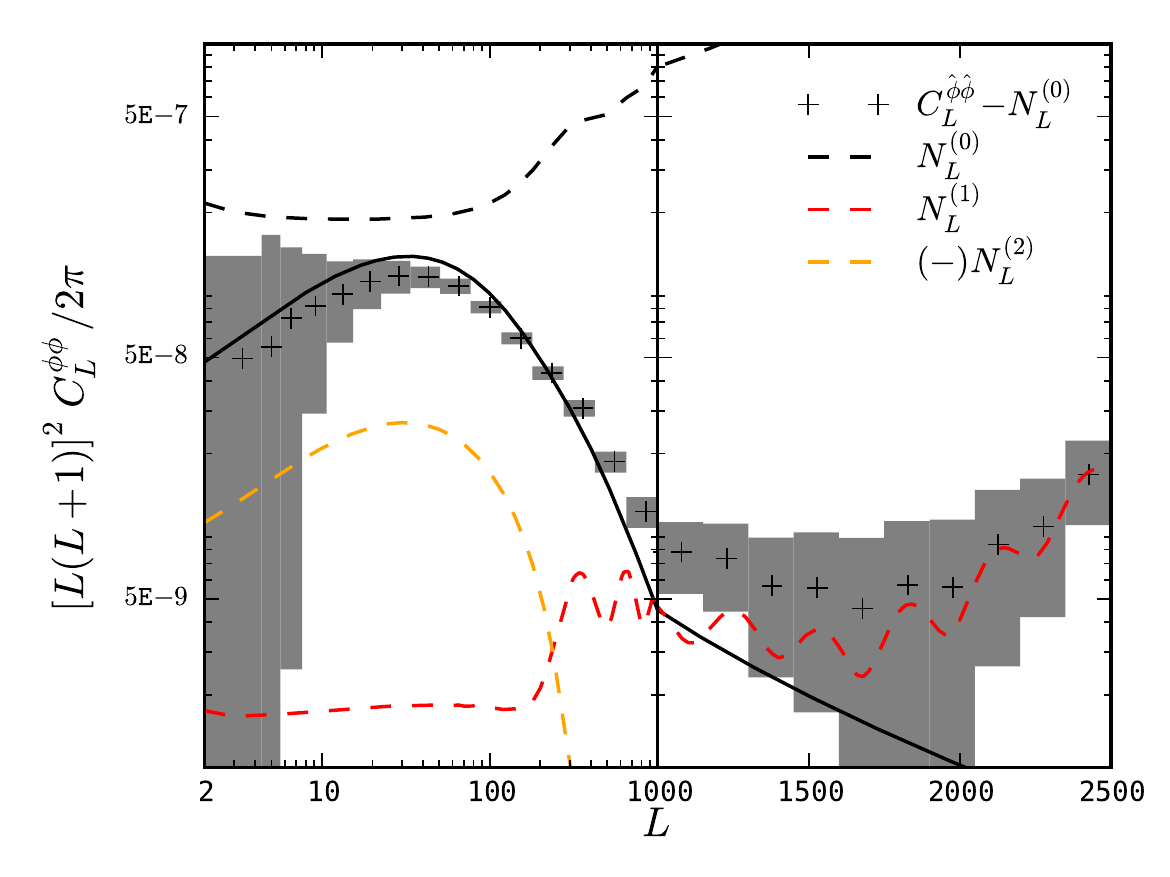}
\caption{Simulations of lensing (deflection) power spectrum recovery $C^{\hat{\phi} \hat{\phi}}_L$. The  ($+$) symbols mark the average power estimate for 1000 simulations, and the gray boxed regions indicate the standard deviation for one realization (with binning over the multipoles corresponding to the width of the box). The black solid line is the input lensing spectrum and the dashed lines are analytical calculations of the bias terms
$N_L^{(0)}$ (black), $N_{L}^{(1)}$
(red) and $N_{L}^{(2)}$ (orange). These are derived in Sec.~\ref{section.psbiases}.}\label{fig.naivespectra}
\end{center}
\end{figure}

In Fig.~\ref{fig.naivespectra} we show simulated reconstructions of the lensing power spectrum. We can see that the $N_{\ell}^{(0)}$ noise bias is the dominant source of error in the reconstruction. The noise bias in the \emph{deflection}
power spectrum $L(L+1)C_L^{\phi\phi}$ is approximately scale-invariant
for $L < 100$. We can understand this behaviour by noting that the
dominant contribution to $N_{L}^{(0)}$ is from chance correlations between
modes near the resolution limit of the experiment. For low $L$, we can therefore
approximate $L \ll \ell_1$, $\ell_2$ in Eq.~(\ref{eqn:gaussianbias}).
Provided $L$ is small compared to the acoustic scale, we can Taylor expand
the $C_{\ell_2}$ power spectra in $\gwt{1}{2}(L)$ about $\ell_1$ to give
\begin{equation}
g_{\el_1\el_2}^2(L)\clexptp{\el_1} \clexptp{\el_2} \approx
\frac{\twolpomul{\el_1 \el_2 L}^2}{16\pi}
\ThreeJ{\el_1}{L}{\el_2}{0}{0}{0}^2 \left(\frac{\cttu{1}}{\clexptp{\el_1}}\right)^2 \left(\llponemul{L}^2 + \frac{(\llponemul{\el_2}^2-\llponemul{\el_1}^2)^2}{4
\llponemul{\el_1}^2} \frac{d\ln \cttu{1}}{d\ln \el_1} \right)^2 .
\end{equation}
Summing over $\el_2$, the leading-order terms for $\el_1 \gg L$ give
\begin{equation}
\sum_{\el_2} g_{\el_1\el_2}^2(L)\clexptp{\el_1} \clexptp{\el_2} \approx
\frac{\twolpomul{\el_1 L}^2\llponemul{L}^4}{16\pi}
\left(\frac{\cttu{1}}{\clexptp{\el_1}}\right)^2 \left[1+
\frac{d\ln \cttu{1}}{d\ln \el_1} + \frac{3}{8} \left(\frac{d\ln \cttu{1}}{d\ln \el_1}\right)^2 \right] .
\label{eq:approxsum}
\end{equation}
Substituting into Eq.~(\ref{eqn:gaussianbias}) and summing over $\ell_1$ gives
$[L(L+1)]^2 N_L^{(0)}/(2\pi) = 2\times 10^{-7}$ in good agreement with
the large-scale plateau in Fig.~\ref{fig.naivespectra}.
The rough scaling with sensitivity and resolution is 
\begin{equation}
[L(L+1)]^2 N_L^{(0)}/(2\pi) \sim 1 / \ell_{\text{max}}^2 , 
\end{equation}
where, here, $\ell_{\text{max}}$ is an effective maximum multipole below which
the measurement of the CMB power spectrum is cosmic-variance limited.
Note that $\ell_{\text{max}}^2$ defines the effective number
of modes and that the noise power scales inversely with this number.

While $N_L^{(0)}$ gives the dominant bias, higher-order terms do have
significant effects. This is illustrated more clearly in Fig.~\ref{fig.biases}, where we plot $\cppest{L} - N_L^{(0)} - \cppp{L} $.
We therefore now set out to investigate the non-Gaussian terms which arise from the connected part of the 4-point function --- $\fourpt_{C}$. To calculate these terms, it is useful to have a system for encoding the constraints which symmetry places on their form. The formalism for doing this has been worked out by \cite{HuAng}, and we briefly review the notation here. 

Statistical isotropy demands that the connected 4-point function function takes the form
\begin{equation}
\fourpt_{C} =
        \sum_{LM} 
                       \ThreeJ{\el_1}{\el_2}{L}
                          {m_1}{m_2}{-M} 
                       \ThreeJ{\el_3}{\el_4}{L}
                          {m_3}{m_4}{M}  (-1)^M 
                        \prim{T}(L)  \,.
\label{eqn:triform}
\end{equation}
The term $\prim{T}(L)$ is known as the trispectrum. 
Inserting this template for the 4-pt function into Eq.~\eqref{eqn:cppestmean} leads to the simplified expression
\begin{equation}
{\langle \cppest{L} \rangle}_{C} = \left( \frac{A_L^2}{\twolpomul{L}^4} \right) \sfourlsnobar g_{\el_{1}\el_{2}}(L) g_{\el_{3}\el_{4}}(L)\prim{T}(L).
\label{eq.purd-i-estdest}
\end{equation}
The trispectrum contains additional symmetries due to our ability to reorder $\el_1 \cdots \el_4$. These may be explicitly encoded by taking
\begin{equation}
T^{\el_1 \el_2}_{\el_3 \el_4}(L) = P^{\el_1 \el_2}_{\el_3 \el_4}(L) + \twolpomul{L}^2 \sum_{L'} 
   \Bigg[ (-1)^{\el_2+\el_3} \wsixj{\el_1}{\el_2}{L}{\el_4}{\el_3}{L'} P^{\el_1 \el_3}_{\el_2 \el_4}(L')  
+         (-1)^{L+L'}        \wsixj{\el_1}{\el_2}{L}{\el_3}{\el_4}{L'} P^{\el_1 \el_4}_{\el_3 \el_2}(L') \Bigg]\,.
\label{eq:sym1}
\end{equation}
There are eight remaining symmetries in the partially-reduced trispectrum representation $\prim{P}(L)$ --- reorderings within the pairings and exchange of upper and lower indices. We may again explicitly symmetrize with respect to pair reorderings via the reduced trispectrum $\trispect$, such that
\begin{eqnarray}
P^{\el_{1} \el_{2}}_{\el_{3} \el_{4}} &=&
\trispect^{\el_{1} \el_{2}}_{\el_{3}\el_{4}}
+ (-1)^{\Sigma_{U}}
\trispect^{\el_{2} \el_{1}}_{\el_{3}\el_{4}}
+ (-1)^{\Sigma_{L}}
\trispect^{\el_{1} \el_{2}}_{\el_{4}\el_{3}}
+ (-1)^{\Sigma_{U} + \Sigma_{L}}
\trispect^{\el_{2} \el_{1}}_{\el_{4}\el_{3}}\,,
\label{eqn:reducedtrispect}
\end{eqnarray}
where $\Sigma_U = \el_1\!+\!\el_2\!+\!L$ and $\Sigma_L=\el_3\!+\!\el_4\!+\!L$. Finally, we may satisfy the remaining upper/lower symmetry with the fully-reduced trispectrum $\prim{\fullred}$ such that
\begin{equation}
\prim{\trispect} = \frac{1}{2} \left[ \prim{\fullred} + \fullred^{\el_{3} \el_{4}}_{\el_{1}\el_{2}} \right] \,.
\label{eqn:fully_reduced_trispectrum}
\end{equation}
The quantity $\prim{\fullred}(L)$ is an arbitrary function of its arguments, and its use enforces all of the symmetry requirements of the trispectrum. These considerations greatly simplify the computation of the lensed trispectrum from the expansion of Eq.~\eqref{eqn:lexphar}, as they make it straightforward to differentiate between terms which are related by symmetries, and terms which are genuinely independent.

In terms of the fully-reduced trispectrum, we can write
\begin{equation}
\fourpt_{C} =
    \frac{1}{2}    \sum_{LM} 
                       \ThreeJ{\el_1}{\el_2}{L}
                          {m_1}{m_2}{-M} 
                       \ThreeJ{\el_3}{\el_4}{L}
                          {m_3}{m_4}{M}  (-1)^M 
                        \prim{\fullred}(L)  + \text{perms.}  \, ,
\end{equation}
which includes a sum over all 24 permutations of $\lbar{1}$, $\lbar{2}$,
$\lbar{3}$ and $\lbar{4}$. This form is particularly convenient for
theoretical calculations of the trispectrum from given physical effects.
It should be noted that fully-reduced trispectrum is only specified up
to the equivalence class that gives the same trispectrum $T^{\el_1 \el_2}_{\el_3 \el_4}(L)$ under the operations in Eqs.~\eqref{eq:sym1}--\eqref{eqn:fully_reduced_trispectrum}. Expanding the trispectrum in Eq.~(\ref{eq.purd-i-estdest}) in terms of
the fully-reduced trispectrum and using the symmetry of the weights
$g_{\el_1\el_2}(L)$ and that they enforce e.g.\ $\el_1+\el_2+L = \text{even}$,
we have
\begin{eqnarray}
{\langle \cppest{L} \rangle}_{C} &=& \left( \frac{4A_L^2}{\twolpomul{L}^4} \right) \sfourlsnobar g_{\el_{1}\el_{2}}(L) g_{\el_{3}\el_{4}}(L)\Biggl[
\prim{\fullred}(L) \nonumber \\
&&\mbox{}+ \twolpomul{L}^2 \sum_{L'} \wsixj{\el_1}{\el_2}{L}{\el_4}{\el_3}{L'}
\left((-1)^{\el_2+\el_3}\fullred^{\el_1 \el_3}_{\el_2 \el_4}(L')
+ (-1)^{L+L'} \fullred^{\el_1\el_3}_{\el_4\el_2}(L')\right) \Biggr].
\label{eq.purd-i-estdest-alt}
\end{eqnarray}
With this considerable array of notation behind us, we are now in a position to calculate the effects of non-Gaussian couplings on the value of $\langle \cppest{L} \rangle$.

\subsection{$\order(\phi^2)$ terms}
\label{subsection:2ndorderterms}

At second order in $\phi$, the connected trispectrum term is due entirely to correlations of the form $\langle \delta\telmn{1} \delta\telmn{2} \telmn{3} \telmn{4} \rangle$. It has the fully-reduced form \cite{HuAng}
\begin{eqnarray}
\prim{\fullred}(L) &=& C_{L}^{\phi\phi} 
\cttu{2} \cttu{4} F_{\el_{1}L\el_{2}}
F_{\el_{3}L\el_{4}} \,.
\label{eqn:lenstrispect}
\end{eqnarray}
Inserting the primary $(\prim{P})$ form of this term into Eq.~\eqref{eq.purd-i-estdest} (or using Eq.~(\ref{eq.purd-i-estdest-alt}) directly) results in
\begin{equation}
{\langle \cppest{L} \rangle}_{C} = \cppp{L} \left( \frac{A_L}{\twolpomul{L}^2}
\sum_{\el_1 \el_2} g_{\el_1\el_2}(L) f_{\el_1 L \el_2}\right)^2 \qquad
\mbox{(primary term)} .
\end{equation}
For any choice of weights, this reduces to $\cppp{L}$ which it was our intention to reconstruct, provided that the estimator is normalized so that the
map-level reconstruction is unbiased, i.e.\ Eq.~(\ref{eq:normalization})
is satisfied. The normalization requires knowledge of the unlensed
CMB power spectrum and any error in the fiducial spectrum used leads to
a multiplicative bias in the reconstructed power.
Inserting the
other pairings into Eq.~\eqref{eq.purd-i-estdest}, on the other hand, results in the first-order power spectrum bias $N_L^{(1)}$:
\begin{equation}
N^{(1)}_L = 2 \left( \frac{A_L}{\twolpomul{L}} \right)^2 \sum_{\el_1 \el_2 \el_3 \el_4 L'} (-1)^{\el_{2}+\el_{3}} \gwt{1}{2}(L) \gwt{3}{4}(L) \wsixj{\el_1}{\el_2}{L}{\el_4}{\el_3}{L'}C^{\phi\phi}_{L'} \lft{\el_1}{L'}{\el_3} \lft{\el_2}{L'}{\el_4}.
\label{eqn:2ndorderbias}
\end{equation}
This $N^{(1)}_L$ term is the all-sky analogue of the $N^{(1)}_{\Theta\Theta,\Theta\Theta}(L)$ noise bias of Kesden, Cooray, and Kamionkowski~\cite{KesCooKam} (KCK
hereafter; see also Appendix~\ref{app.flatskytrispectrum}).
In practice, we find that the flat-sky result gives a good description of the biases found in our simulations. This is a reasonable result as the magnitude of the $N_{L}^{(1)}$ bias is non-negligible only at large multipoles and here flat-sky calculations converge with their harmonic counterparts. For the remainder of this paper, we will calculate $N_L^{(1)}$ using the flat-sky expression, with all-sky power spectra as input.

In Fig.~\ref{fig.naivespectra} we plot the $N^{(1)}_L$ bias. At low multipoles it is an insignificant contribution, but at $L\!>\!250$ it produces an $\order(10\%)$ effect. At $L\!>\!1000$ it begins to dominate over the $\cppp{L}$ term which it was our intention to reconstruct. We show in Fig.~\ref{fig.biases} that
at high multipoles the $N^{(1)}_L$ bias provides a good fit to the excess
power seen in our simulation results.

There are a number of ways to account for this behaviour. Hu showed that for lensing, the primary $(\prim{P})$ component of the trispectrum dominates over the other couplings when min$(\el_1,\el_2,\el_3,\el_4) \gg L$~\cite{HuAng}
The primary terms represent flattened quadrilaterals with the short diagonal of length $L$ supported by lensing, the power of which is strongly peaked at low
multipoles.
A heuristic (but frequently useful) way of rephrasing this argument in a way which is specific to lens reconstruction comes from considering the effect of the weights $\gwt{1}{2}(L)$. For $\el_1 \gg L$, the weights are only non-zero
for $\el_2$ in a narrow range of width $2L$ centered on $\el_1$ (due to the
triangle constraint).
The $N_L^{(1)}$ term imposes couplings between two weight terms, and so the region of  $(\el_1, \el_2, \el_3, \el_4)$ space over which it can accumulate signal is correspondingly suppressed. In general, at low-$L$ the dominant trispectrum terms which bias $\cppest{L}$ are therefore those which factor maximally under the weights. At high-$L$, which terms will dominate is less clear. The eventual dominance of the $N_L^{(1)}$ bias was explained by KCK as the interference of lensing modes at $L'<L$ on the reconstruction. It appears that in this case the cross-coupling nature of the non-primary terms which suppresses their contribution at low-$L$ allows them to dominate later on. As $C_L^{\phi\phi}$ is very red, the coupling of the $N^{(1)}_L$ term to $C_L^{\phi\phi}$ power at lower-$L$ allows it to increase relative to the primary term.

\subsection{$\order(\phi^{4})$ terms}
\label{subsection.4thorderterms}

At second order in the power spectrum, there are several fundamental groupings of $\delta^{n}\Theta$ and unlensed terms in the lensing expansion which contribute to the fully-reduced trispectrum. These are all contained in four fully-reduced trispectra:
\begin{eqnarray}
\overset{{\delta\Theta \delta\Theta \delta\Theta \delta\Theta}}{ \prim{\fullred} (L) } &=& \frac{1}{2}\sum_{\el_a \el_b \el_c \el_d} \Biggl( (-1)^{L+\el_2+\el_3}
\twolpomul{L}^2 \wsixj{\el_1}{\el_2}{L}{\el_a}{\el_b}{\el_c} \wsixj{\el_3}{\el_4}{L}{\el_a}{\el_b}{\el_d} \nonumber \\
&&\mbox{}\hspace{0.15\textwidth}\times \bfft{1}{b}{c} \bfft{2}{a}{c} \bfft{3}{b}{d} \bfft{4}{a}{d}
\cttu{a} \cttu{b} \cpp{c} \cpp{d} \Biggr) \nonumber \\
\overset{\delta^2\Theta \delta^2\Theta \tilde{\Theta} \tilde{\Theta}}{ \prim{\fullred} (L) } &=& (-1)^{L+\el_1+\el_2} 2 \twolpomul{L}^2 \twolpomul{\el_1 \el_2 \el_3 \el_4} \llponemul{\el_2 \el_4} \nonumber\\
&&\mbox{}\hspace{0.1\textwidth}\times \sum^\prime_{\el_a \el_b} \left( \frac{\twolpomul{\el_a \el_b} \llponemul{\el_a \el_b}}{16\pi} \right)^2 \cttu{2} \cttu{4} \cpp{a} \cpp{b} \jfish{\el_1}{\el_2}{\el_a}{\el_b}{L} \jfish{\el_3}{\el_4}{\el_a}{\el_b}{L}  \nonumber \\
\overset{ {\delta^2\Theta \delta\Theta \delta\Theta \tilde{\Theta}} }{ \prim{\fullred}(L) } &=& - \cppp{L} R \left(  \llponemul{\el_1}^2 \cttu{1} \bfftL{2}{1} \right) \left(  \cttu{3} \bfftL{4}{3} \right) \nonumber \\
&&\mbox{}+ \cppp{L} \left[ \bfftL{2}{1} \sum_{\el_a \el_b} \cttu{a}\cpp{b} (\bfft{1}{a}{b})^2 \left( \frac{\llponemul{\el_1}^2 + \llponemul{\el_a}^2 - \llponemul{\el_b}^2}{\twolpomul{\el_1}^2 \llponemul{\el_1}^2} \right) \right] \left( \cttu{3} \bfftL{4}{3}  \right) \nonumber \\
&&\mbox{}+ \frac{1}{8\pi}\sum^\prime_{\el_a \el_b \el_c} \Biggl( (-1)^{\el_1+\el_2+\el_c}\cttu{3}\cttu{c}\cpp{a}\cpp{b} \bfft{1}{b}{c} \bfft{2}{a}{c} \nonumber\\
&&\mbox{} \hspace{0.1\textwidth}\times \twolpomul{L}^2\twolpomul{\el_3\el_4\el_a\el_b}\llponemul{\el_3\el_a\el_c} \jfish{\el_4}{\el_3}{\el_b}{\el_a}{L}
\wsixj{\el_1}{\el_2}{L}{\el_a}{\el_b}{\el_c} \Biggr)\nonumber \\
\overset{ {\delta^3\Theta \delta\Theta \tilde{\Theta} \tilde{\Theta}} }{ \prim{\fullred} (L) } &=& - \cppp{L} R \left[  \left(\llponemul{\el_1}^2-2/3\right) \cttu{1} \bfftL{2}{1} \right] \left(  \cttu{3} \bfftL{4}{3} \right) .
\label{eqn:4thorderterms}
\end{eqnarray}
Here, the value $R$ is half of the mean-square deflection
\begin{equation}
R = \frac{1}{2} \sum_{\el_1} \frac{ (\llponemul{\el_1} \twolpomul{\el_1})^2 }{4\pi} \cpp{1} \, ,
\label{eqn:Rall}
\end{equation}
and the $\jfish{\el_1}{\el_2}{\el_3}{\el_4}{L}$ function is given in Appendix~\ref{app.expansionterms}. The
summation $\sum^\prime_{\el_a \el_b}$ in the second grouping is restricted
to $\el_a$ and $\el_b$ such that $\el_a + \el_b + \el_1 + \el_2 = \text{even}$
and $\el_a + \el_b + \el_3 + \el_4 = \text{even}$, thus enforcing the
parity constraint $\el_1 + \el_2 + \el_3 + \el_4 = \text{even}$.
Similarly, the summation $\sum^\prime_{\el_a\el_b\el_c}$ in the third grouping
is restricted such that $\el_3 + \el_4 + \el_a + \el_b=\text{even}$.

\begin{figure}
\begin{center}
\includegraphics[width=5in]{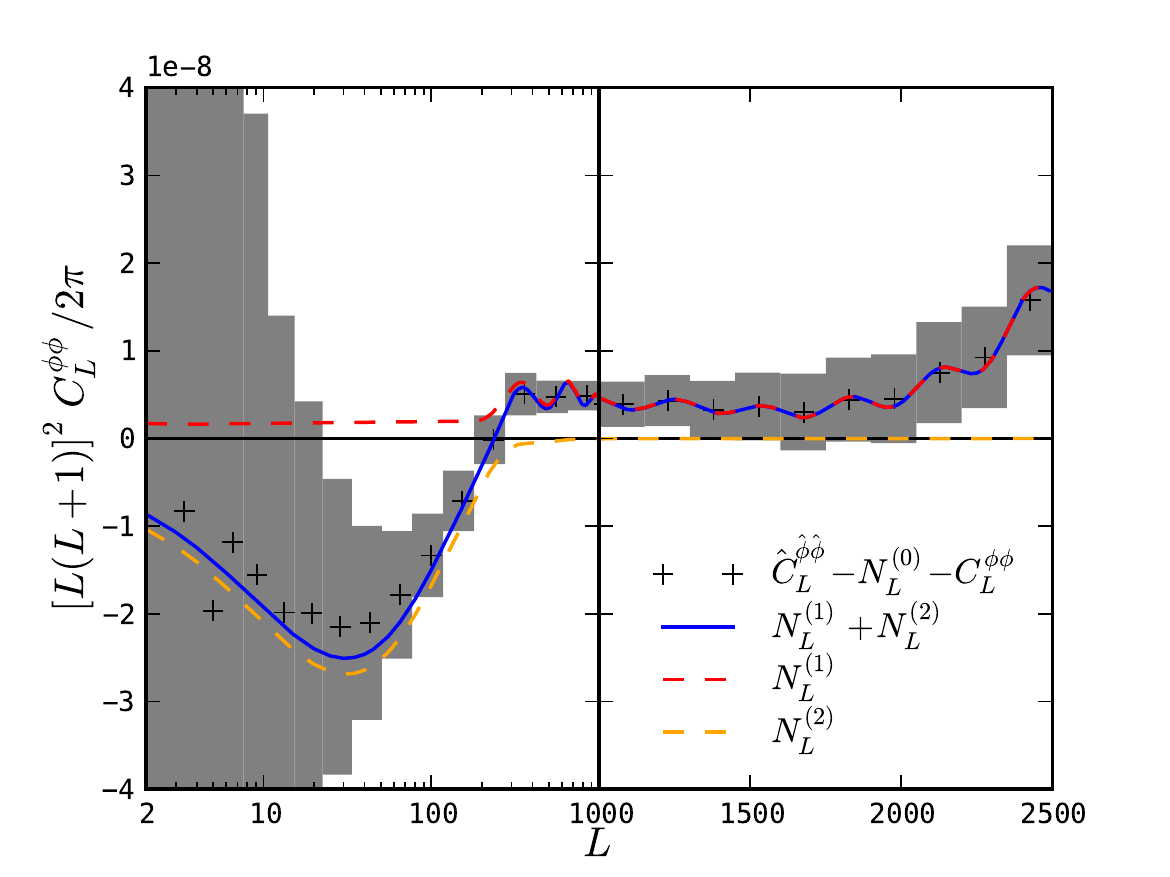}
\caption{Biases in $\cppest{L}$.
The points are the average of 1000 simulations and the boxes give the measured error in the reconstruction for one realization (with binning over the multipoles corresponding to the width of the box). The biases calculated in Sec.~\ref{section.psbiases} are also shown: $N_L^{(1)}$ dotted (red); $N_L^{(2)}$
dashed (orange); and their sum in solid (blue). These calculated biases
provide a good fit to the observed power in simulations.}\label{fig.biases}
\end{center}
\end{figure}

Many of these terms are difficult to calculate numerically. Fortunately, the dominant subset \textit{is} calculable:
\begin{equation}
\overset{\rm{dom}}{ \prim{\fullred} } (L) = \cppp{L}
\left( \cttu{3} \bfftL{4}{3}  \right)\bfftL{2}{1} 
\Bigg[\sum_{\el_a \el_b} \cttu{a} \cpp{b} (\bfft{1}{a}{b})^2 \left( \frac{\llponemul{\el_1}^2 + \llponemul{\el_a}^2 - \llponemul{\el_b}^2}{\twolpomul{\el_1}^2 \llponemul{\el_1}^2} \right) 
- 2R (\llponemul{\el_1}^2 -1/3)\cttu{1} \Bigg] .
\label{eqn:domofterms}
\end{equation}
The choice of the terms in Eq.~\eqref{eqn:domofterms} is motivated by the discussion of the previous subsection, in that these terms are the subset of Eq.~\eqref{eqn:4thorderterms} which maximally factor under the weights for the primary trispectrum coupling. Calculating the effect of $\overset{\text{dom}}{ \prim{\fullred} } (L)$ through the primary trispectrum coupling results in a low-$\ell$ bias at ${\cal O}(\phi^4)$:
\begin{eqnarray}
N_{L}^{(2)} &\approx & 2 \cppp{L} \left(\frac{A_{L}^2}{\twolpomul{L}^4}\right)
\sum_{\el_1\el_2} \gwt{1}{2}(L) \bfftL{2}{1}
\bigg[
 \sum_{\el_{a}\;\el_{b}} \cttu{a} \cpp{b} (\bfft{1}{a}{b})^2 \left( \frac{\llponemul{\el_1}^2 + \llponemul{\el_a}^2 - \llponemul{\el_b}^2}{\twolpomul{\el_1}^2 \llponemul{\el_1}^2} \right) \nonumber \\
&&\mbox{} \hspace{0.3\textwidth}
-2R(\llponemul{\el_1}^2-1/3) \cttu{1} \Bigg]
\left(\sum_{\el_1\el_2} \gwt{3}{4}(L) f_{\el_3 L \el_4} \right).
\label{eqn:nl2bias}
\end{eqnarray}
We plot these dominant terms to the $N_L^{(2)}$ bias in
Figs.~\ref{fig.naivespectra} and \ref{fig.biases}.
Overall, they contribute a \textit{negative} bias to the reconstructed power spectrum which is $\approx 15\%$ of $C_L^{\phi\phi}$ for the concordance cosmology which we use. They provide a good fit to the negative bias seen in our simulations
for $L < 200$.

The dominant terms that we have retained in Eq.~(\ref{eqn:nl2bias})
share a common origin: they are due to correlations of the gradient-order potential reconstruction of $\hat{\phi}_{LM}$ with higher order terms. Equivalent terms arise, for example, if one calculates $\langle C_L^{\phi \hat{\phi}} \rangle$ --- the cross-spectrum between the input and reconstructed lensing potentials. 
If we write the reconstructed $\hat{\phi}_{LM}$ as\footnote{We are grateful
to Antony Lewis for suggesting this approach.}
\begin{equation}
\hat{\phi}_{LM} = n^{(0)}_{LM} + \phi_{LM} + n^{(1)}_{LM} + n^{(2)}_{LM} +
n^{(3)}_{LM} + \cdots,
\label{eq:AML1}
\end{equation}
where, for example, $n^{(2)}_{LM}$ is the statistical noise in the reconstruction that is second order in $\phi$ and arises from $\tilde{\Theta}\delta^2 \Theta$
and $\delta \Theta \delta \Theta$ terms, then the $\mathcal{O}(C_L^{\phi\phi})$
contribution to
$\langle C_L^{\phi \hat{\phi}} \rangle$ is from $\phi_{LM}$ only (since
$\langle n^{(1)}_{LM} \rangle_{\text{CMB}}=0$ by construction) and
the $\mathcal{O}[(C_L^{\phi\phi})^2]$ contribution is from $n^{(3)}_{LM}$.
Correct to fourth order in $\phi$, we find
\begin{equation}
\langle C_L^{\phi \hat{\phi}} \rangle = \cppp{L} + \frac{1}{2} \overset{\text{dom}}{N_L^{(2)}} ,
\label{eq:AML2}
\end{equation}
where $\overset{\text{dom}}{N_L^{(2)}} = 2 \langle \phi_{LM}^* n^{(3)}_{LM} \rangle$
is given by the right-hand side of Eq.~(\ref{eqn:nl2bias}) with one factor
of the normalization eliminated with Eq.~(\ref{eq:normalization}).

We can develop a useful approximation to $N_L^{(2)}$ on large scales by noting
that the sum over $\el_1$ and $\el_2$ is dominated by small-scale modes near
the resolution limit of the experiment. Since $C_L^{\phi\phi}$ is so red,
$\el_b \ll \el_a \approx \el_1$ in the remaining summation and we can approximate
\begin{equation}
\sum_{\el_a \el_b} \cttu{a} \cpp{b} (\bfftp{L}{\el_a}{\el_b})^2 \frac{[\llponemul{L}^2 + \llponemul{\el_a} - \llponemul{\el_b}^2]}{\llponemul{L}^2 \twolpomul{L}^2} \approx 2 \sum_{\el_a \el_b} \cttu{a} \cpp{b} \frac{(\bfftp{L}{\el_a}{\el_b})^2}{\twolpomul{L}^2} .
\label{eq:approxNtwo}
\end{equation}
Noting further that the second-order result for the change in the temperature
power spectrum due to lensing is~\cite{HuHar}
\begin{equation}
\Delta\cttlp{L} = \cttlp{L} - \cttup{L} \approx -\llponemul{L}^2 R \,\cttup{L} + \sum_{\el_a \el_b} \cttu{a} \cpp{b} \frac{(\bfftp{L}{\el_a}{\el_b})^2}{\twolpomul{L}^2},
\label{eqn:lensed_ps}
\end{equation}
we can approximate
\begin{equation}
N_{L}^{(2)} \approx  4 \cppp{L} \left(\frac{A_{L}}{\twolpomul{L}^2}\right)^2
\left(\sum_{\el_3\el_4} g_{\el_3\el_4}(L)f_{\el_3 L \el_4} \right)
\sum_{\el_1\el_2} \gwt{1}{2}(L) \bfftL{2}{1} \Delta\cttlp{\ell_1} .
\label{eq:NtwoDeltaCl}
\end{equation}
The \emph{fractional} $N_{L}^{(2)}$ bias is therefore largely determined by the
difference between the lensed and unlensed power spectra.
We can use similar methods to those which gave Eq.~(\ref{eq:approxsum})
to approximate $N_{L}^{(2)}$ further at low $L$:
\begin{equation}
N_L^{(2)}\approx  \cppp{L} \frac{A_L \llponemul{L}^4}{4\pi}
\sum_{\ell_1} \frac{\twolpomul{\el_1}^2}{\left(\clexptp{\el_1}\right)^2} \left[
\cttu{1} \Delta \cttlp{\ell_1} + \frac{1}{2}\frac{d}{d\ln\ell_1}
\left( \Delta \cttlp{\ell_1}\cttu{1}\right)
+ \frac{3}{8} \frac{d\cttu{1}}{d\ln \ell_1}\frac{d\Delta\cttlp{\ell_1}}{d\ln \ell_1} \right] ,
\label{eq:approxNtwobias}
\end{equation}
where we have used Eq.~(\ref{eq:normalization}) for the normalization.
The only $L$ dependence is in the prefactor, but Eqs.~(\ref{eqn:gaussianbias})
and~(\ref{eq:approxsum}) show that $A_L \llponemul{L}^4 = \text{const.}$ at low
$L$ so the $N^{(2)}_L$ bias is a constant fractional bias there. Numerically,
the dominant contribution to Eq.~(\ref{eq:approxNtwobias}) is from the final
term and the bias evaluates to $-20\%$, in reasonable agreement with the
full evaluation of Eq.~(\ref{eqn:nl2bias})
reported in Fig.~\ref{fig.biases}. The bias become less important for surveys
with lower resolution: for an experiment limited to a maximum
multipole $\el_{\text{max}} = 750$, we find the bias reduces to $-4\%$.

We have seen that the dominant terms in the $\mathcal{O}(\phi^4)$ trispectrum
provide a good explanation of the low-$\ell$ bias seen in our simulation results.
Figure~\ref{fig.biases} provides empirical evidence that higher-order terms
do not bias the reconstructed power further by more than $\sim 3\%$. This seems
plausible given Eq.~(\ref{eq:NtwoDeltaCl}). Replacing $\Delta \cttlp{\ell_1}$
by its non-perturbative equivalent (see e.g.~Ref.~\cite{ChaLewCorr}), the
change in $\Delta \cttlp{\ell_1}$ is around 20\% at $\ell_1 \sim 2000$. To the
extent that this replacement mimics the non-perturbative generalization of the
low-$L$ bias (see Appendix~\ref{app.flatskytrispectrum} for further justification), we should expect higher-order terms to produce a fractional bias
$\sim 0.15\times 0.2 = 0.03$.

\section{Power Spectrum Covariance}
\label{section.pscovariance}
We now proceed to investigate the covariance of our power spectrum estimates. A similar calculation has been performed on the flat sky by \cite{KesCooKam}, however the full-sky calculation is somewhat cleaner. 

We are interested in the covariance of the reconstructed power spectrum which is given by
\begin{equation}
\cov( \cppest{L}, \cppest{L'} ) = \langle \cppest{L} \cppest{L'} \rangle - \langle \cppest{L} \rangle \langle \cppest{L'} \rangle.
\label{eqn:psestcov}
\end{equation}
The cross term,
\begin{multline}
\left< \cppest{L} \cppest{L'} \right> = {\left( \frac{A_L A_{L'}}{\twolpomul{L L'}} \right)}^2 \sum_{M=-L}^{L} \sum_{M'=-L'}^{L'} \sum_{\lbar{1}\lbar{2}\lbar{3}\lbar{4}} \sum_{\lbar{5}\lbar{6}\lbar{7}\lbar{8}} (-1)^{M+M'}\\
\times \ThreeJ{\el_1}{\el_2}{L}{\m_1}{\m_2}{M} \ThreeJ{\el_3}{\el_4}{L}{\m_3}{\m_4}{-M} \ThreeJ{\el_5}{\el_6}{L'}{\m_5}{\m_6}{M'} \ThreeJ{\el_7}{\el_8}{L'}{\m_7}{\m_8}{-M'} \\
\times \gwt{1}{2}(L) \gwt{3}{4}(L) \gwt{5}{6}(L') \gwt{7}{8}(L') \eightpt ,
\label{eqn:pscovcrossterm}
\end{multline}
requires us to calculate the 8-point lensed correlation function. Since the
lensed CMB has zero mean, and we are ignoring the low-$\el$ $C_\el^{\tilde{\Theta}\phi}$ correlation and hence all odd connected correlations, the 8-point function
can be expressed in terms of the connected 2-, 4-, 6- and 8-point functions.
The latter two of these are at least $\mathcal{O}(\phi{}^4)$~\cite{2002PhRvD..66h3007K} and so we neglect them on the grounds that they should
be sub-dominant.

We first consider those contributions to the 8-point function that involve the
trispectrum. There are 70 terms involving products of two trispectra and a
further 210 involving the product of a trispectrum with two 2-point functions.
Guided by our experience with the power spectrum, it seems likely that the dominant couplings are those which factor most under the weights. This motivates us
to consider only those terms with trispectra which couple terms between
two weights, e.g. $\langle \T_{\lbar{1}}\T_{\lbar{2}}\T_{\lbar{3}}\T_{\lbar{4}} \rangle_{C}$. There are three such terms involving a product of trispectra and
18 involving a single trispectrum. It is convenient to re-express these in
terms of the full 4-point functions and their Gaussian (disconnected) parts,
so their contribution to the 8-point function is
\begin{multline}
\langle \T_{\lbar{1}}\T_{\lbar{2}}\T_{\lbar{3}}\T_{\lbar{4}} \rangle
\langle \T_{\lbar{5}}\T_{\lbar{6}}\T_{\lbar{7}}\T_{\lbar{8}} \rangle
- \left(\langle \T_{\lbar{1}}\T_{\lbar{2}}\rangle \langle\T_{\lbar{3}}\T_{\lbar{4}} \rangle + \text{2 perms}\right)\\
\times \left(\langle \T_{\lbar{5}}\T_{\lbar{6}}\rangle \langle\T_{\lbar{7}}\T_{\lbar{8}} \rangle + \text{2 perms}\right) + (34\leftrightarrow 56)
+ (34 \leftrightarrow 78) .
\label{eq:fourpoint}
\end{multline}
The 4-point terms make simple contributions to Eq.~(\ref{eqn:pscovcrossterm}):
\begin{eqnarray}
\fourpta{1}{2}{3}{4} \fourpta{5}{6}{7}{8} &\rightarrow& \langle \cppest{L} \rangle \langle \cppest{L'} \rangle \nonumber \\
\fourpta{1}{2}{5}{6} \fourpta{3}{4}{7}{8} &\rightarrow& \delta_{LL'}\langle \cppest{L} \rangle \langle \cppest{L'} \rangle / (2L+1)  \nonumber \\
\fourpta{1}{2}{7}{8} \fourpta{3}{4}{5}{6} &\rightarrow& \delta_{LL'} \langle \cppest{L} \rangle \langle \cppest{L'} \rangle / (2L+1) .
\label{eqn:dompsvar}
\end{eqnarray} 
For $L=L'$ this leads to a dominant variance of 
\begin{equation}
\var( \cppest{L} ) = \frac{2}{2L+1} \langle \cppest{L} \rangle^2 ,
\label{eqn:psvarterm}
\end{equation}
which agrees with the results of \cite{KesCooKam} on the flat sky. It also
justifies the assumption of \cite{HuMap} that the statistical noise in the $\hat{\phi}$ reconstruction behaves predominantly like Gaussian noise giving the usual cosmic-variance result for the power spectrum variance. This is a slightly different version of cosmic variance than we usually deal with, however, in that the terms in $\cppest{L}$ arise from correlations between a large number of multipoles.

\begin{figure}
\begin{center}
\includegraphics[width=5in]{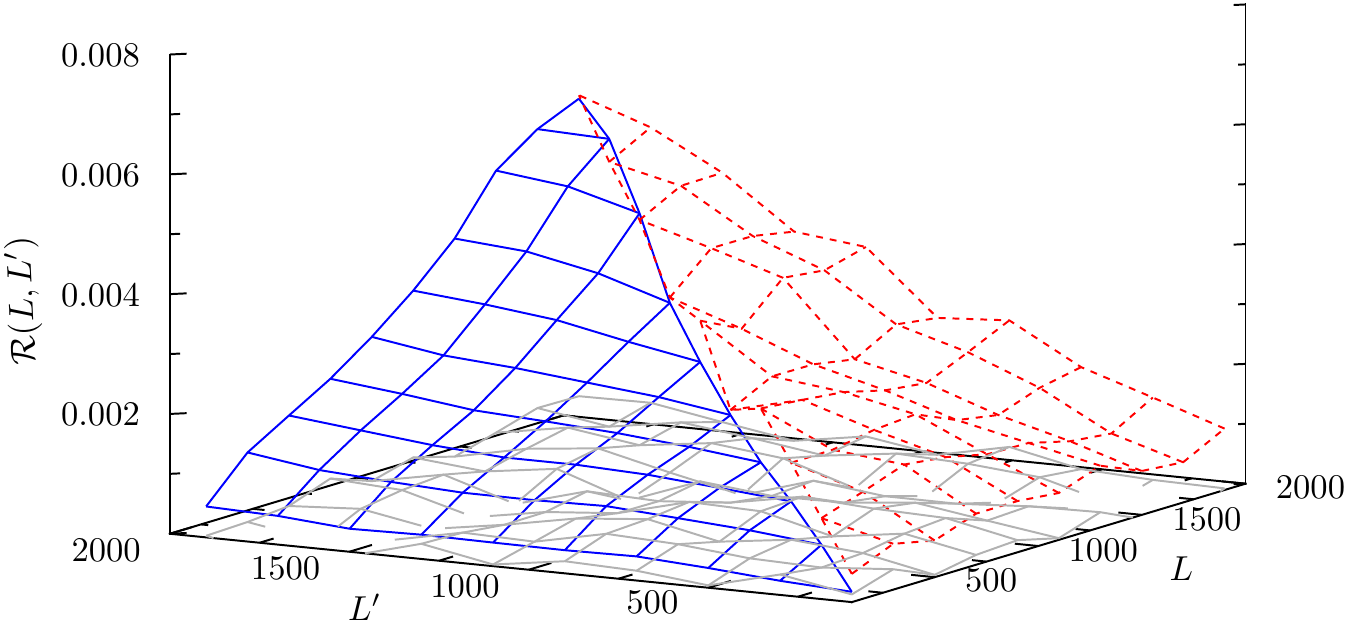}
\caption{Covariance $\rcov(L,L')$ calculated approximately using Eq.~\eqref{eqn:pscovtermAresult} (solid blue) and estimated from 1000 simulations using the standard (constant) $N^{(0)}_L$ bias term of Eq.~\eqref{eqn:gaussianbias} (dashed red) and the realization-dependent modified expression of Eq.~\eqref{eqn:gaussianbiasmod} (solid gray). }
\label{fig.psrcov}
\end{center}
\end{figure}

We now consider the disconnected part of the 8-point function and the
remaining 2-point terms in Eq.~(\ref{eq:fourpoint}). As we shall see,
these generate a covariance between the power spectrum estimates.
The disconnected part of the 8-point function involves 105 terms but
45 of these are zero (for $L$ and $L' \neq 0$) because they correlate one or more pairs of $\ThetaL$ which are jointly weighted, which isolates $\langle\hat{\phi}_{LM}\rangle=0$ when averaged over realizations of large scale structure.
The remaining 60 non-zero terms are characterized by four fundamental pairings,
with terms within each pairing giving equal contributions to the covariance
by symmetry:
\begin{subequations}
\begin{align}
\contraction{}{\T_{\lbar{1}}}{}{\T_{\lbar{3}}}
\contraction{\T_{\lbar{1}}\T_{\lbar{3}}}{\T_{\lbar{2}}}{}{\T_{\lbar{4}}}
\contraction{\T_{\lbar{1}}\T_{\lbar{3}}\T_{\lbar{2}}\T_{\lbar{4}} \ \mid \ }
{\T_{\lbar{5}}}{}{\T_{\lbar{7}}}
\contraction{\T_{\lbar{1}}\T_{\lbar{3}}\T_{\lbar{2}}\T_{\lbar{4}} \ \mid \ 
\T_{\lbar{5}}\T_{\lbar{7}}}{\T_{\lbar{6}}}{}{\T_{\lbar{8}}}
\T_{\lbar{1}}\T_{\lbar{3}}\T_{\lbar{2}}\T_{\lbar{4}} \ \mid \ 
\T_{\lbar{5}}\T_{\lbar{7}}\T_{\lbar{6}}\T_{\lbar{8}} 
\quad \quad \quad \quad & \text{(4 terms)} 
\label{eqn:pscovtermA} \\
\contraction[4ex]{}{\T_{\lbar{1}}}{\T_{\lbar{2}}\T_{\lbar{3}}\T_{\lbar{4}}\ \mid \ \T_{\lbar{5}}\T_{\lbar{6}} \T_{\lbar{7}}}{\T_{\lbar{8}}}
\contraction[3ex]{\T_{\lbar{1}}}{\T_{\lbar{2}}}{\T_{\lbar{3}}\T_{\lbar{4}}\ \mid \ \T_{\lbar{5}}\T_{\lbar{6}}}{\T_{\lbar{7}}}
\contraction[2ex]{\T_{\lbar{1}}\T_{\lbar{2}}}{\T_{\lbar{3}}}{\T_{\lbar{4}}\ \mid \ \T_{\lbar{5}}}{\T_{\lbar{6}}}
\contraction{\T_{\lbar{1}}\T_{\lbar{2}}\T_{\lbar{3}}}{\T_{\lbar{4}}}{\ \mid \ }{\T_{\lbar{5}}}
\T_{\lbar{1}}\T_{\lbar{2}}\T_{\lbar{3}}\T_{\lbar{4}}\ \mid \ \T_{\lbar{5}}\T_{\lbar{6}} \T_{\lbar{7}}\T_{\lbar{8}}
\quad \quad \quad \quad & \text{(8 terms)} \label{eqn:pscovtermC1} \\
\contraction{}{\T_{\lbar{1}}}{}{\T_{\lbar{3}}}
\contraction[2ex]{\T_{\lbar{1}}\T_{\lbar{3}}}{\T_{\lbar{2}}}{\T_{\lbar{4}}
\ \mid \ \T_{\lbar{7}}}{\T_{\lbar{5}}}
\contraction{\T_{\lbar{1}}\T_{\lbar{3}}\T_{\lbar{2}}}{\T_{\lbar{4}}}
{\ \mid \ }{\T_{\lbar{7}}}
\contraction{\T_{\lbar{1}}\T_{\lbar{3}}\T_{\lbar{2}}\T_{\lbar{4}}
\ \mid \ \T_{\lbar{7}} \T_{\lbar{5}}}{\T_{\lbar{6}}}{}{\T_{\lbar{8}}}
\T_{\lbar{1}}\T_{\lbar{3}}\T_{\lbar{2}}\T_{\lbar{4}}
\ \mid \ \T_{\lbar{7}} \T_{\lbar{5}}\T_{\lbar{6}}\T_{\lbar{8}}
\quad \quad \quad \quad & \text{(32 terms)} \label{eqn:pscovtermB} \\
\contraction[4ex]{}{\T_{\lbar{1}}}{\T_{\lbar{2}}\T_{\lbar{3}}\T_{\lbar{4}}\ \mid \ \T_{\lbar{7}}\T_{\lbar{5}} \T_{\lbar{6}}}{\T_{\lbar{8}}}
\contraction[3ex]{\T_{\lbar{1}}}{\T_{\lbar{2}}}{\T_{\lbar{3}}\T_{\lbar{4}}\ \mid \ \T_{\lbar{7}}\T_{\lbar{5}}}{\T_{\lbar{6}}}
\contraction[2ex]{\T_{\lbar{1}}\T_{\lbar{2}}}{\T_{\lbar{3}}}{\T_{\lbar{4}}\ \mid \ \T_{\lbar{7}}}{\T_{\lbar{5}}}
\contraction{\T_{\lbar{1}}\T_{\lbar{2}}\T_{\lbar{3}}}{\T_{\lbar{4}}}{\ \mid \ }{\T_{\lbar{7}}}
\T_{\lbar{1}}\T_{\lbar{2}}\T_{\lbar{3}}\T_{\lbar{4}}\ \mid \ \T_{\lbar{7}}\T_{\lbar{5}} \T_{\lbar{6}}\T_{\lbar{8}}
\quad \quad \quad \quad & \text{(16 terms)} . \label{eqn:pscovtermC2}
\end{align}
\end{subequations}
The first and second pairings cancel with the 12 non-zero 2-point terms in Eq.~(\ref{eq:fourpoint}). The third pairing results in a covariance of
\begin{equation}
\langle \cppest{L} \cppest{L'} \rangle_{a} = 32 {\left( \frac{A_L A_{L'}}{\twolpomul{L L'}} \right)}^2 \sum_{\el_1} \frac{1}{\twolpomul{\el_1}^2} \left(\clexptp{\el_1} \right)^2 \left[ \sum_{\el_2} \gwt{1}{2}^2(L) \clexptp{\el_2} \right] \left[ \sum_{\el_{3}} \gwt{1}{3}^2(L') \clexptp{\el_3} \right].
\label{eqn:pscovtermAresult}
\end{equation}
The final pairing is more tightly coupled, and results in
\begin{multline}
\langle \cppest{L} \cppest{L'} \rangle_{b} = 16 {\left( \frac{A_L A_{L'}}{\twolpomul{L L'}} \right)}^2 \sum_{\el_1\el_2\el_3\el_4} \Biggl(
(-1)^{L+L'} \wsixj{L}{\el_1}{\el_2}{L'}{\el_3}{\el_4}  \clexptp{\el_1}\clexptp{\el_2}\clexptp{\el_3}\clexptp{\el_4} \\
\times \gwt{1}{2}(L) \gwt{3}{4}(L) \gwt{1}{4}(L') \gwt{2}{3}(L') \Biggr).
\label{eqn:pscovtermBresult}
\end{multline}
We expect that the terms of Eq.~\eqref{eqn:pscovtermAresult} should be dominant as they factor most under the weights, and so from here on we consider them only.
In Fig.~\ref{fig.psrcov} we plot the correlation matrix

\begin{equation}
\rcov(L,L') = \frac{\cov(\cppest{L},\cppest{L'})}{\sqrt{\var(\cppest{L})\var(\cppest{L'})}}.
\end{equation}
The agreement between the measured and theoretically-approximated correlations is excellent. Using the approximation in Eq.~(\ref{eq:approxsum}), the
correlation matrix for low $L$ and $L'$ should scale as
$\rcov(L,L') \sim \sqrt{LL'}/l_{\text{max}}^2$ in agreement with the
arguments in Ref.~\cite{KesCooKam}. Note that the correlations decrease
as the resolution of the experiment increases.
While the shape of our numerical
covariance agrees with~\cite{KesCooKam}, the magnitude we find is at least one order larger.
The correlations are at a level of $<\!1\%$, although we note that this correlation is for unbinned spectra. With such broad correlations, binning increases the correlation roughly in proportion to the bin width. The binned spectra in Fig.~\ref{fig.naivespectra}, for example, have correlations of ${\cal O}(1\%)$ for $\el < 1000$, where logarithmic binning is used, and of ${\cal O}(10\%)$ for $\el > 1000$, where the bins are rather wider. In the discussion which follows, however, we will show how these correlations may be almost entirely eliminated in practice.

\section{Discussion}

We have identified three complications with the usual lensing estimator:
\begin{enumerate}
\item{the ${\cal O}(\phi^2)$ $N_L^{(1)}$ excess power bias;}
\item{the ${\cal O}(\phi^4)$ $N_L^{(2)}$ power suppression; and}
\item{a small intrinsic covariance between the estimated $C_L^{\hat{\phi}\hat{\phi}}$}.
\end{enumerate}
We will discuss the implications of each of these terms and their remediation in turn.

\begin{figure}
\begin{center}
\includegraphics[width=5in]{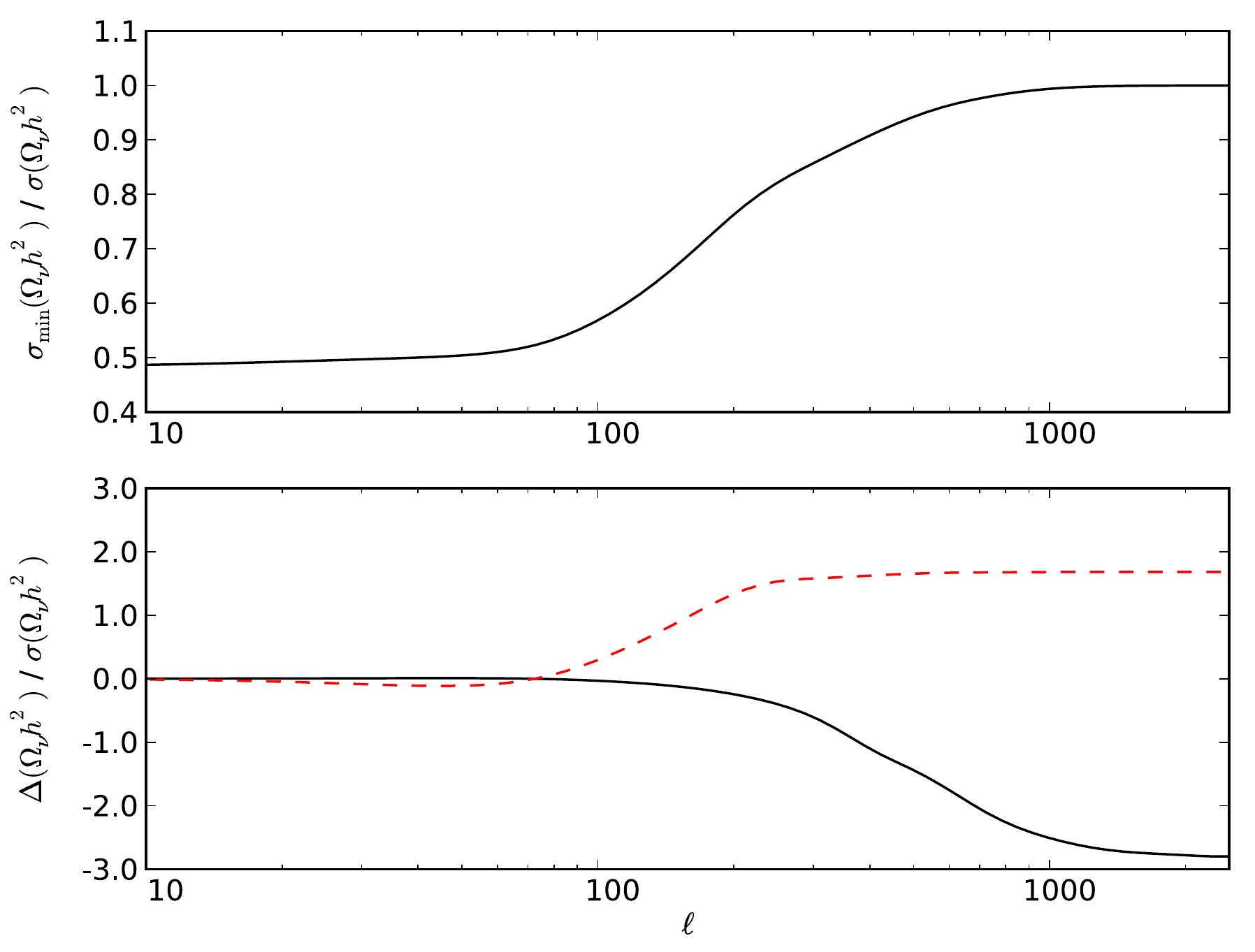}
\caption{Relationship between $C_L^{\phi\phi}$ reconstruction and parameter constraints for \Planck\!. 
Top panel: constrains on $\Omega_{\nu}h^2$ assuming that the $C_L^{\phi\phi}$ reconstruction is discarded for $L>\ell$, divided by the constraint if the full reconstruction were included. Inclusion of lens reconstruction improves constraints on $\Omega_{\nu}h^2$ by a factor of two, consistent with e.g. \cite{LesPerPasPia}.
Bottom panel: bias in the determination of $\Omega_{\nu}h^2$, as a fraction of the random
error, induced by the $N^{(1)}_L$ (black solid) and $N^{(2)}_L$ (red dashed) biases, assuming that the biases have been completely removed for $L>\ell$. }\label{fig.fisher}
\end{center}
\end{figure}

We begin by investigating the relevance of the bias terms for cosmological parameter determination with a Fisher matrix approach. Our implementation follows closely that of \cite{Perotto}, and we refer the interested reader there for details. 
The incorporation of $C_L^{\phi\phi}$ reconstruction into a parameter analysis allows one to produce dramatically improved constraints on the `dark' parameters which affect the late-time evolution of the Universe \cite{KapKnoSon}. With \Planck for example, a factor of two improvement over what is achievable without
lensing reconstruction is forecasted for the determination of the (summed)
neutrino mass in simple models, and rather more in models with additional parameters
such as dynamical dark energy~\cite{LesPerPasPia, Perotto}. We begin by asking which region of the $C_L^{\phi\phi}$ power spectrum this determination is made from. We use the fiducial cosmology given in Sec.~\ref{section.simulations}, but take $\Omega_{\nu}h^2=0.006$ (i.e. $\sum_\nu m_\nu = 0.6 \, \text{eV}$).
This is large enough that it would be detected with significance by \Planck\!, which is important to prevent the hard $\Omega_{\nu}h^2 \ge 0$ prior from corrupting the Fisher analysis. In the top panel of Fig.~\ref{fig.fisher} we plot the relative constraints on $\Omega_{\nu} h^2$ for our simplified version of \Planck\!, assuming that the reconstructed $C_L^{\phi \phi}$ power spectrum is ignored above some value of $L$. We can see from this figure that for \Planck\!, most of the constraining power of lensing on neutrino masses comes from the multipole range $100\!<\!L\!<\!700$, which will receive large contributions from both bias terms.

We now consider (pessimistically) the effect of the $N^{(1)}_L$ and $N^{(2)}_L$ biases if completely uncorrected. We can relate biases in the `observed' lensing power spectrum to parameter biases by perturbing the Fisher-approximated likelihood, finding that
\begin{equation}
\Delta \theta_{i} = F^{-1}_{ij} \sum_{AB,CD} [\partial C^{AB}/\partial\theta_j]  [\Delta \hat{C}^{CD}] [{\rm Cov}^{-1}]_{AB,CD},
\label{eqn:fisher_biases}
\end{equation}
where $\theta_{i}$ indexes a cosmological parameter, $F_{ij}$ is the Fisher matrix, $\Delta\hat{C}^{CD}$ is the power spectrum bias and $[{\rm Cov}]$ is the covariance matrix of the measured spectra. The labels $A$, $B$, $C$ and $D$ run over the observable fields $T$, $E$ and $\phi$.
In the lower panel of Fig.~\ref{fig.fisher} we assume that the bias is uncorrected for $L<\el$ and perfectly removed for $L>\el$. We can see that neglect of either bias leads to $1\sigma$ errors in determination of the neutrino energy density. The $N_L^{(2)}$ bias suppresses the reconstructed power, particularly at low-$\el$, and this partly mimics the effect of massive neutrinos, leading to an overestimate of $\Omega_{\nu}h^2$. This is particularly worrisome if we are placing an upper limit on the neutrino mass, as neglect of the $N_L^{(2)}$ bias would lead to a spurious (albeit marginal) detection of massive neutrinos. We conclude that an accurate treatment of both biases must be made in order to obtain accurate parameter constraints, and so now turn to ways of mitigating these.

For the $N_{L}^{(1)}$ bias, the suggestion for removal has been to perform iterative estimation and subtraction \cite{KesCooKam, AmbValWhi}.
An additional approach, which is perhaps conceptually more straightforward, is to view $N_{L}^{(1)}$ as a normalization effect which results in an estimator which is a convolution over the true lensing power spectrum rather than a direct estimate. The kernel of this convolution can be calculated and inverted to produce an unbiased estimate of $C_L^{\phi\phi}$. We note that this kernel depends on the true unlensed CMB power spectrum. In practice, our uncertainty in this quantity therefore determines the extent to which the $N_{L}^{(1)}$ bias may be treated. Due to the large noise with which upcoming experiments will reconstruct the lensing potential, both of these procedures are likely to be unstable unless we can take the shape of the lensing power to be characterized by some small set of numbers (such as the cosmological parameters of interest).

Experimental realism introduces additional difficulties, however. In particular, we note that for a realistic experiment with inhomogeneous sky coverage, the normalization of a quadratic lensing estimator must generally be determined with simulations~\cite{SmiZahDor}. To treat the $N_L^{(1)}$ bias using off-diagonal entries in a normalization matrix would require that we be able to Monte-Carlo these elements --- a difficult task. In this case, the $N_L^{(1)}$ bias will most likely need to be absorbed into the normalization. Suppose, for example, that the normalization is determined with lensed realizations of the CMB using some fiducial $C_{L}^{\phi\phi}$ spectrum, as in~\cite{SmiZahDor}. At each $L$, a normalization $A_L$ is determined from the ratio of the Monte-Carlo average of the
reconstructed power (after correction for the $N_L^{(0)}$ noise)
to the fiducial power there.
Assuming that the \emph{shape} of the fiducial $C_{L}^{\phi\phi}$ matches the underlying lensing power, this results in an unbiased estimator at the power spectrum level, albeit with slightly greater variance than would exist in the absence of the $N_L^{(1)}$ bias. The quality of our assumptions about the underlying $C_{L}^{\phi\phi}$ shape would then need to be quantified and folded into the systematic-error estimate.

Treatment of the $N_L^{(2)}$ bias presents further difficulties due to its non-linearity in $C_L^{\phi\phi}$. It is particularly troublesome in the case where the normalization is determined by Monte-Carlo simulations, as care must be made to determine the normalization over a small range of $C_L^{\phi\phi}$ power, to ensure that the estimator response is approximately linear. The $N_L^{(2)}$ bias would then appear as an offset which could be subtracted in a fiducial model with simulations using
the fiducial $C_L^{\phi\phi}$ power.
This would clearly be a cumbersome procedure. We find, however, that a small modification of the optimal estimator can effectively remove this bias prior to normalization. Our starting point is the approximate expression for the
$N^{(2)}_L$ bias in Eq.~(\ref{eq:approxNtwo}). In terms of this, we have
\begin{eqnarray}
\langle C_L^{\hat{\phi}\hat{\phi}} \rangle &=& C_L^{\phi\phi} \left(\frac{A_L}{%
\twolpomul{L}^2}\right)^2 \left[\left(\sum_{\el_1 \el_2} g_{\el_1 \el_2}(L)
f_{\el_1 L \el_2}\right)^2 + 2 \sum_{\el_1 \el_2} g_{\el_1 \el_2}(L)
f_{\el_1 L \el_2} \sum_{\el_3 \el_4} g_{\el_3 \el_4}(L)
f_{\el_3 L \el_4}^\Delta \right] + N_L^{(0)} + N_L^{(1)} \nonumber \\
&\approx& C_L^{\phi\phi} \left(\frac{A_L}{\twolpomul{L}^2}\right)^2
\left[\sum_{\el_1 \el_2} g_{\el_1 \el_2}(L)\left(f_{\el_1 L \el_2}
+ f^\Delta_{\el_1 L \el_2}\right)\right]^2 + N_L^{(0)} + N_L^{(1)} ,
\end{eqnarray}
where $f_{\el_1 L \el_2}^\Delta = \Delta C_{\el_2}^{\Theta\Theta} f_{\el_1 L \el_2}
+ \Delta C_{\el_1}^{\Theta\Theta} f_{\el_2 L \el_1}$ and, recall, $\Delta C_\el^{\Theta\Theta}$ is the change in the CMB power spectrum due to lensing. We now see that if we
modify the weights to
\begin{equation}
\bar{g}_{\el_1 \el_2}(L) = \frac{f_{\el_1 L \el_2} - f^\Delta_{\el_1 L \el_2}}{
2\cttlp{\el_{1}, {\rm expt}}\cttlp{\el_{2}, {\rm expt}}} ,
\label{eqn:nl2impwghts}
\end{equation}
where $f^\Delta_{\el_1 L \el_2}$ is computed for some fiducial $C_L^{\phi\phi}$,
the quadratic part of the estimator response to power $C_L^{\phi\phi}$
becomes linear times the difference between the true lensed
CMB power and that in the fiducial model. The estimator normalization, for
an ideal survey, is still given by Eq.~(\ref{eqn:gaussianbias}) using the
original weights. In practice, the normalization could be obtained
by Monte-Carlo with simulations using the fiducial $C_L^{\phi\phi}$ and
the modified weights in the reconstruction.
The cost of this modification is an increase in the estimator variance of $\sim\!15\%$ for $L<300$, however this is most likely justified given the improved bias properties of the estimator. Additionally, this technique makes clear that our ability to debias the estimator is determined by our understanding of $C_L^{\phi\phi}$ only through the lensed and unlensed CMB temperature power spectra, a simplicity which one would not necessarily expect for a higher-order bias. This insight makes the assessment of systematic errors due to uncertainties in cosmology much more straightforward.

Another, more straightforward method of treating the bias is motivated by the recent work of Lewis et. al. \cite{Lewis:2011fk}, 
who point out that to a good degree of approximation, the response of the lensed CMB covariance to changes in the lensing 
potential is determined by lensed power spectra. This is in contrast to the first-order formula used to derive the quadratic estimators 
of Okamoto and Hu \cite{OkHu}, which contains an unlensed spectrum. Incorporating this insight into the quadratic derivation one would obtain the 
same estimator, but with lensed rather than unlensed spectra in the filtering of Eq.~\eqref{eq:gonetwo}. Because these appear 
twice in the estimator normalization (Eq.~\eqref{eq:normalization}), this is in fact equivalent at $\order(\phi^{4})$ to the correction described above. 
This method of debiasing is even simpler to implement than the one proposed above.

Finally, we consider the covariance of the lensing power reconstruction. The loss of information which this covariance represents increases the effective error bars of the reconstruction, although we have not quantified this here as we find that an improved, internally-calibrated estimator does a good job of reducing this covariance. Similar ideas were put forward by Dvorkin and Smith in the context of optical-depth reconstruction~\cite{DvoSmi}.
For lens reconstruction, the modified estimator for the power spectrum is as follows. Instead of removing the $N_L^{(0)}$ bias by direct subtraction, we
subtract $2\hat{N}_L^{(0)} - N_L^{(0)}$, where $\hat{N}_L^{(0)}$ involves the
observed total power spectrum in our realization of the universe,
$\hat{C}_{\el, {\rm expt}}^{\ThetaL \ThetaL}$:
\begin{equation}
\hat{N}_L^{(0)} = \frac{A_L^2}{\twolpomul{L}^2} \sum_{\el_1\,\el_2} [\gwt{1}{2}(L)]^2 [2 C_{\el_1, {\rm expt}}^{\ThetaL \ThetaL} \hat{C}_{\el_2, {\rm expt}}^{\ThetaL \ThetaL} ] .
\label{eqn:gaussianbiasmod}
\end{equation}
By construction, $\hat{N}_L^{(0)}$ is quadratic in the observed temperature
and has $\langle \hat{N}_L^{(0)} \rangle = N_L^{(0)}$. 
Subtracting $2\hat{N}_L^{(0)} - N_L^{(0)}$
is very successful in reducing the intrinsic covariance of the estimator, while
preserving the expectation value of the power.\footnote{%
This modification to the estimator is also what one finds if one constructs
a quartic estimator in the observed temperature for $C_L^{\phi\phi}$ based
on an approximate maximization of the likelihood for the lensed temperature
(truncated at trispectrum order).}
This is because the dominant terms of
$\text{Cov}(\cppest{L} , \hat{N}_{L'}^{(0)}) + \text{Cov}(\cppest{L'} , \hat{N}_{L}^{(0)})$ are identical to those of Eq.~\eqref{eqn:pscovtermAresult}, and
the dominant terms of $\text{Cov}(\hat{N}_{L}^{(0)},\hat{N}_{L'}^{(0)})$ are
$1/4$ of those in Eq.~\eqref{eqn:pscovtermAresult}. Their combination
therefore cancels the dominant contribution to the off-diagonal covariance
from $\langle C_L^{\hat{\phi}\hat{\phi}} C_L^{\hat{\phi}\hat{\phi}} \rangle_a$.
With this improved estimator, we find that the covariance is reduced to a level which is unmeasurable in our simulations $(<0.05\%)$. For an experiment with only partial sky coverage, $\hat{N}_L^{(0)}$ can be determined by simulations for
each of which the quadratic $\phi$ reconstruction is performed by
correlating independent Gaussian CMB maps drawn from $\hat{C}_{\el, {\rm expt}}^{\ThetaL \ThetaL}$ and $C_{\el, {\rm expt}}^{\ThetaL \ThetaL}$ respectively.

As we move toward experimental realism, the number of quantities which are determined by Monte-Carlo becomes increasingly worrisome, although some sanity checks are available. The accuracy of the $N_L^{(0)}$ subtraction, for example, may be assessed by a jackknife test in which one reconstructs the lensing power using two $\phi$ estimates: one which correlates only even multipoles, and one which correlates only odd multipoles. For parity-symmetric sky coverage, this results in an estimate of $C_L^{\phi\phi}$ which is free from the $N_L^{(0)}$ bias, providing a useful consistency check.

\section{Conclusions}
We have thoroughly investigated the behaviour of the optimal quadratic lensing estimator, both analytically and with simulations. We have discovered a new bias in the reconstructed power spectrum, and presented the lensed CMB trispectrum at ${\cal O}(\phi^4)$ to explain it. This bias has a physical interpretation as an anti-correlation between the first-order reconstruction of $\phi$ and higher-order terms, and so it will also be relevant for cross-correlation studies. We suggest that the following estimator provides a good basis for the reconstruction of $C_L^{\phi\phi}$:
\begin{equation}
\hat{C}_L^{\phi\phi} = {\cal A}_L^{L'} \bar{C}_{L'}^{\hat{\phi}\hat{\phi}} - 2\hat{N}_{L}^{(0)} + N_L^{(0)} ,
\end{equation}
where $\bar{C}_{L'}^{\hat{\phi}\hat{\phi}}$ denotes the usual optimal estimator of Eq.~\eqref{eqn:cppestdef} with a slight modification of the weights, given by Eq.~\eqref{eqn:nl2impwghts}. This estimator has a slightly larger variance than the standard optimal estimator at low $L$, however it is effectively free from contamination by higher-order lensing terms and has negligible covariance. Uncertainty in the underlying cosmology leads to potentially imperfect removal of the biasing terms, however we find analytically that this possibility is completely characterized by our understanding of the lensed and unlensed temperature spectra.

In the near future, CMB lensing will make the transition from detection to precision science, and the concerns which we have addressed will be increasingly important. The reconstruction of the large-scale lensing potential is a demanding task, however the new window which it will open onto the contents of the Universe will most certainly provide a worthwhile view.

\section{Acknowledgments}
The work carried out in this paper made use of the \healpix \cite{GorskHeal} package
for pixelization.
We thank Antony Lewis and Kendrick Smith for helpful discussions; in particular
Antony contributed several important insights to the discussion around
Eq.~(\ref{eq:AML1}). PB thanks the Agence Nationale de la Recherche grant ANR-05-BLAN-0289-01 for support.

\appendix
\section{Simplification of Expansion Terms}
\label{app.expansionterms}

We wish to calculate the connected part of the trispectrum to ${\mathcal O}(\phi^4)$, which requires a third-order expansion of the lensing effect. In harmonic space, these terms are given by
\begin{equation}
\begin{array}{lll}
\delta\telm &=& \sum_{\lbar{1}\lbar{2}} \phi_{\lbar{1}} \ttelmn{2} \itn \\
\delta^2\telm &=& \frac{1}{2} \sum_{\lbar{1}\lbar{2}\lbar{3}} \phi_{\lbar{1}} \phi_{\lbar{2}} \ttelmn{3} \jtn \\
\delta^3\telm &=& \frac{1}{6} \sum_{\lbar{1}\lbar{2}\lbar{3}\lbar{4}} \phi_{\lbar{1}} \phi_{\lbar{2}} \phi_{\lbar{3}} \ttelmn{4} \ktn ,
\end{array}
\end{equation}
where
\begin{equation}
\begin{array}{lll}
\itn &=& \intsky{ {\Ylmn}^{*} (\gYlm{i}{1}) \nabla^{i}\Ylm{2} }\\
\jtn &=& \intsky{ {\Ylmn}^{*} (\gYlm{i}{1}) (\gYlm{j}{2}) \nabla^{i}\nabla^{j}\Ylm{3}} \\
\ktn &=& \intsky{ {\Ylmn}^{*} (\gYlm{i}{1}) (\gYlm{j}{2}) (\gYlm{k}{3}) \nabla^{i}\nabla^{j}\nabla^{k}\Ylm{4}} .
\end{array}
\label{eq.expansionintegrals}
\end{equation}
We may simplify the covariant derivatives in Eq.~\eqref{eq.expansionintegrals} by exploiting their relationship with the spin-raising and lowering operators \cite{LewChaTur}. Their application thus results in the generation of spin-$s$ spherical harmonics, which are defined by the application of the spin-raising and lowering operators to the ordinary (spin-0) harmonics~\cite{NewPenNotes, GolMacSpinS}. Following the derivation in \cite{OkHu}, it is straightforward to show that
\begin{equation}
\nabla^{k} \left( \Yslmn  \vc{e}_{+}^{(n_{+})} \vc{e}_{-}^{(n_{-})} \right)
= -\frac{1}{2} \vc{e}_{+}^{(n_{+})} \vc{e}_{-}^{(n_{-})} \Bigl( \sqrt{(\el-s)(\el+s+1)} \left[ \Yslm{s+1}{\el}{m} \right] \vc{e}_{-}^{k}
-\sqrt{(\el+s)(\el-s+1)} \left[ \Yslm{s-1}{\el}{m} \right] \vc{e}_{+}^{k}  \Bigr), 
\label{eqn:yslmderiv}
\end{equation}
where $n_{-} - n_{+}=s$, the (null) spin basis is given in terms of the
unit polar and azimuthal basis vectors, $\epm \equiv (\etheta \pm i \ephi)$,
and we have defined $\vc{e}_{\pm}^{(n)} = \vc{e}_{\pm}^{i_1} \vc{e}_{\pm}^{i_2} \cdots \vc{e}_{\pm}^{i_n}$. Repeated applications of $\nabla$ reduce to applications of Eq.~(\ref{eqn:yslmderiv}). 
We now proceed to simplify the expansion terms individually.
\subsection{$\itn$}
The simplification of $\itn$ has been given in the general case by \cite{HuAng}. Expanding the gradients of the spherical harmonics and then contracting indices gives 
\begin{equation}
\itn = -\frac{1}{2} \llponemul{\el_1 \el_2} \intsky{ {\Ylmn}^* \left( \Yslmp{1}{1} \Yslmp{-1}{2} + \Yslmp{-1}{1} \Yslmp{1}{2} \right) } .
\end{equation}
This vanishes by parity unless $\ell + \el_1 + \el_2 = \text{even}$.
Now we may use an identity for integration over three spherical harmonics \cite{VarQuant},
\begin{equation}
\intsky{} \Yslm{s}{l}{m}{}^* \Yslm{s_1}{l_1}{m_1} \Yslm{s_2}{l_2}{m_2}
= (-1)^{m+s}
 \frac{\twolpomul{\el \el_1 \el_2}}{\sqrt{4\pi}} \ThreeJ {\el}{\el_1}{\el_2}{s}{-s_1}{-s_2}
 \ThreeJ{\el}{\el_1}{\el_2}{-m}{m_1}{m_2}\,,
\label{eq.intthreey}
\end{equation}
and the following result from the recursion relations between the $3j$ symbols~\cite{VarQuant}:
\begin{equation}
\ThreeJ{\el_1}{\el_2}{\el_3}{1}{-1}{0} = \frac{1}{2} \ThreeJ{\el_1}{\el_2}{\el_3}{0}{0}{0} \left( \frac{\llponemul{\el_3}^2 - \llponemul{\el_1}^2 - \llponemul{\el_2}^2}{\llponemul{\el_1 \el_2}} \right) \quad \quad (\el_1\!+\!\el_2\!+\!\el_3 \ \text{even}) ,
\label{eq.recurone}
\end{equation}
to show that
\begin{equation}
\itn = (-1)^{m} \ThreeJ{\el}{\el_1}{\el_2}{-\m}{\m_1}{\m_2} \bfftp{\el}{\el_1}{\el_2},
\end{equation}
where the parity constraint is enforced by $\bfftp{\el}{\el_1}{\el_2}$.
This simple result for $\itn$
also follows rather more directly by repeatedly integrating by
parts in Eq.~(\ref{eq.expansionintegrals}).

\subsection{$\jtn$}

It follows from parity that $\jtn$ vanishes unless $\ell+\ell_1+\ell_2+\ell_3
= \text{even}$. When this is satisfied, we have
\begin{equation}
\jtn = \frac{\llponemul{\el_1 \el_2 \el_3}}{2}
\sqrt{(\ell_3+2)(\ell_3-1)}
\intsky{}{\Ylmn}^* \! \Yslmp{-1}{1} \! \Yslmp{-1}{2} \! \Yslmp{2}{3}
+ \frac{\llponemul{\el_1 \el_2 \el_3}\llponemul{\el_3}}{2}
\intsky{}{\Ylmn}^* \! \Yslmp{1}{1} \! \Yslmp{-1}{2} \! \Yslmp{0}{3} .
\end{equation}
Thus the gradient integral reduces to an integral over four spin spherical harmonics. We may reduce this to a three-harmonic integral by invoking the Clebsch-Gordan expansion for spin harmonics \cite{HuWhi}
\begin{equation}
\Yslm{\es_1}{\el_1}{\m_1} \Yslm{\es_2}{\el_2}{\m_2} = \frac{\twolpomul{\el_1 \el_2}}{\sqrt{4\pi}}\sum_{LMS} \twolpomul{L} \ThreeJ{\el_1}{\el_2}{L}{-\es_1}{-\es_2}{S} \ThreeJ{\el_1}{\el_2}{L}{\m_1}{\m_2}{-M} \Yslm{S}{L}{M} (-1)^{M-S} .
\end{equation}
Application of the three-harmonic integral identity of Eq.~\eqref{eq.intthreey} then gives
\begin{eqnarray}
\intsky \Yslm{\es_1}{\el_1}{\m_1}\Yslm{\es_2}{\el_2}{\m_2}\Yslm{\es_3}{\el_3}{\m_3}\Yslm{\es_4}{\el_4}{\m_4} &=& \frac{\twolpomul{\el_1 \el_2 \el_3 \el_4}}{4\pi}\sum_{LMS}\twolpomul{L}^2 \left[\ThreeJ{\el_1}{\el_2}{L}{-\es_1}{-\es_2}{-S} \ThreeJ{\el_1}{\el_2}{L}{\m_1}{\m_2}{M} \right. \nonumber \\
&&\mbox{}\times \left. (-1)^{M-S}  \ThreeJ{\el_3}{\el_4}{L}{-\es_3}{-\es_4}{S} \ThreeJ{\el_3}{\el_4}{L}{\m_3}{\m_4}{-M} \right].
\end{eqnarray}
This may be used finally to give two compact expressions for $\jtn$ for
$\ell+\ell_1+\ell_2 +\ell_3= \text{even}$:
\begin{eqnarray}
\jtn &=& (-1)^{m_1}\frac{\twolpomul{\el \el_1 \el_2 \el_3} \llponemul{\el_1 \el_2 \el_3}}{8\pi} \sum_{L} \twolpomul{L}^2 \ThreeJ{\el}{\el_1}{L}{-\m}{\m_1}{M} \ThreeJ{\el_2}{\el_3}{L}{\m_2}{\m_3}{-M} \jhook{\el}{\el_1}{\el_2}{\el_3}{L} \nonumber\\
&=& (-1)^{m_3}\frac{\twolpomul{\el \el_1 \el_2 \el_3} \llponemul{\el_1 \el_2 \el_3}}{8\pi} \sum_{L} \twolpomul{L}^2 \ThreeJ{\el}{\el_3}{L}{-\m}{\m_3}{M} \ThreeJ{\el_1}{\el_2}{L}{\m_1}{\m_2}{-M} \jfish{\el}{\el_3}{\el_1}{\el_2}{L}, 
\end{eqnarray}
where
\begin{eqnarray}
\jhook{\el}{\el_1}{\el_2}{\el_3}{L} &=& - \left[ \ThreeJ{\el}{\el_1}{L}{0}{-1}{1} \ThreeJ{\el_2}{\el_3}{L}{1}{0}{-1} \llponemul{\el_3} +  \ThreeJ{\el}{\el_1}{L}{0}{1}{-1} \ThreeJ{\el_2}{\el_3}{L}{1}{-2}{1} \sqrt{(\el_3 -1)(\el_3 + 2)} \right] \nonumber \\
\jfish{\el}{\el_3}{\el_1}{\el_2}{L} &=&  \left[ \ThreeJ{\el}{\el_3}{L}{0}{0}{0} \ThreeJ{\el_1}{\el_2}{L}{1}{-1}{0} \llponemul{\el_3} + 
  \ThreeJ{\el}{\el_3}{L}{0}{2}{-2} \ThreeJ{\el_1}{\el_2}{L}{-1}{-1}{2} \sqrt{(\el_3 -1)(\el_3 + 2)} \right].
\end{eqnarray}
Note that $\jhook{\el}{\el_1}{\el_2}{\el_3}{L}$ and $\jfish{\el}{\el_3}{\el_1}{\el_2}{L}$ correspond to the two distinct ways of coupling the arguments
of $J$ (it is symmetric on the second and third arguments). They are therefore
related by a 6$j$ recoupling coefficient (for $\el + \el_1 + \el_2 +\el_3 =\text{even}$):
\begin{equation}
\jfish{\el}{\el_3}{\el_1}{\el_2}{L} = \sum_{L'} (-1)^{\el+\el_3+L'}
\twolpomul{L'}^2 \wsixj{\el}{\el_3}{L}{\el_2}{\el_1}{L'}
\jhook{\el}{\el_1}{\el_2}{\el_3}{L'} .
\end{equation}
In our calculations in the main text, we only encounter $\jhookn$ terms with
$(\el+\el_1\!+\!L)$ and $(\el_2\!+\!\el_3\!+\!L)$ even. In this case
we can simplify further by using the additional recursion result
\begin{equation}
\ThreeJ{\el_1}{\el_2}{\el_3}{1}{1}{-2} = \frac{1}{2} \ThreeJ{\el_1}{\el_2}{\el_3}{0}{0}{0} \left[ \frac{\llponemul{\el_1}^2(\llponemul{\el_3}^2 - \llponemul{\el_1}^2 + \llponemul{\el_2}^2) + \llponemul{\el_2}^2 ( \llponemul{\el_3}^2 + \llponemul{\el_1}^2 - \llponemul{\el_2}^2 ) }{\llponemul{\el_1 \el_2 \el_3} \sqrt{(\el_3 - 1)(\el_3+2)}} \right] \, ,
\label{eq.recurtwo}
\end{equation}
which holds only for $(\el_1\!+\!\el_2\!+\!\el_3)$ even. Using Eqs.~(\ref{eq.recurone}) and~(\ref{eq.recurtwo}) we find two simple equivalent forms for $\jhook{\el}{\el_1}{\el_2}{\el_3}{L}$:
\begin{equation}
\jhook{\el}{\el_1}{\el_2}{\el_3}{L} = \frac{4\pi \bfftp{\el}{L}{\el_1}}{(\twolpomul{L}\llponemul{L})^2 \twolpomul{\el \el_1 \el_2 \el_3} \llponemul{\el_1 \el_2 \el_3}}
\times
\left\{
\begin{array}{ll}
\bfftp{L}{\el_2}{\el_3} (\llponemul{\el_3}^2 - \llponemul{\el_2}^2 + \llponemul{L}^2)& \\
&\\
\bfftp{\el_2}{L}{\el_3} (\llponemul{\el_3}^2 + \llponemul{\el_2}^2 - \llponemul{L}^2). & 
\end{array}
\right.
\end{equation}

\subsection{$\ktn$}

Using the tools introduced thus far we may produce a general expression for the $\ktn$ term. In our calculations, however, we only encounter terms in which
$K$ is contracted on its second and third or third and fourth sets of indices.
The first of these is
\begin{equation}
\sum_{\m_1} (-1)^{\m_1} K_{\el \el_1 \el_1 \el_3 \el_4}^{\m \m_1 -\m_1 \m_3 \m_4} =  \intsky{{\Ylmn}^{*} \left[ \sum_{\m_1} \gYlm{i}{1} \nabla_{j} Y_{\el_1}^{\m_1*} \right] (\gYlm{k}{3}) \nabla^{i}\nabla^{j}\nabla^{k} \Ylm{4}}.
\end{equation}
Invoking the result that \cite{HuHar}
\begin{eqnarray}
\sum_m \nabla_i \Ylmn \nabla_j {\Ylmn}^* &=& \frac{(\llponemul{\el} \twolpomul{\el})^2 }{8\pi} g_{ij} ,
\end{eqnarray}
where $g_{ij}$ is the metric on the unit sphere,
and using
\begin{equation}
\nabla^2 \nabla^k \Ylmn = \nabla^k (\nabla^2 + 1) \Ylmn = \left(1-
\llponemul{\el}^2\right) \nabla^k \Ylmn ,
\end{equation}
we find that
\begin{equation}
\sum_{\m_1} (-1)^{\m_1} K_{\el \el_1 \el_1 \el_3 \el_4}^{\m \m_1 -\m_1 \m_3 \m_4} = 
\left(1-\llponemul{\el_4}^2\right)\frac{(\llponemul{\el_1} \twolpomul{\el_1})^2}{8\pi} I_{\ell \ell_3 \ell_4}^{\m \m_3 \m_4} .
\end{equation}
A similar calculation for the contraction of $K$ on its third and
fourth indices gives
\begin{equation}
\sum_{\m_2} (-1)^{\m_2} K_{\el \el_1 \el_2 \el_2 \el_4}^{\m \m_1 \m_2 -\m_2 \m_4} = 
-\frac{(\llponemul{\el_4}\llponemul{\el_2} \twolpomul{\el_2})^2}{8\pi} I_{\ell \ell_1 \ell_4}^{\m \m_1 \m_4} .
\end{equation}

\section{Flat-Sky Trispectrum}
\label{app.flatskytrispectrum}
For small patches of the sky, the curvature of the sphere is negligible and we can represent the CMB in terms of Fourier modes rather than spherical harmonics. Geometrical terms in this flat-sky basis are represented as scalar-products, which frequently makes them easier to work with than the corresponding full-sky expressions. Here we present expressions for the flat-sky trispectrum following the
Fourier conventions of~\cite{LewChaLens}.


The flat-sky trispectrum, $T(\elb_1, \elb_2, \elb_3, \elb_4)$, is defined by
\begin{equation}
\langle \T(\elb_1) \T(\elb_2) \T(\elb_3) \T(\elb_4) \rangle_{C} = (2\pi)^{-2} \delta( \elb_1+\elb_2+\elb_3+\elb_4) T(\elb_1, \elb_2, \elb_3, \elb_4).
\label{eq:appb1}
\end{equation}
The trispectrum is permutation symmetric in all its vector arguments and rotational and parity invariance mean that it depends on five scalar parameters of the quadrilateral formed from its arguments. As with the spherical trispectrum, it is frequently convenient to work with a fully-reduced trispectrum, $\fullred^{(\ell_1 \ell_2)}_{(\ell_3 \ell_4)}(L)$, such that
\begin{equation}
\langle \T(\elb_1) \T(\elb_2) \T(\elb_3) \T(\elb_4) \rangle_{C} =
\frac{1}{2}\int \frac{d^2 \vL}{(2\pi)^2} \delta(\elb_1 + \elb_2 + \vL)\delta(\elb_3 + \elb_4 -\vL) \fullred^{(\ell_1 \ell_2)}_{(\ell_3 \ell_4)}(L) + \text{perms.} ,
\label{eq:appb2}
\end{equation}
where we include the sum over all permutations of $\elb_1$, $\elb_2$, $\elb_3$
and $\elb_4$. The fully-reduced trispectrum is then an arbitrary function of its arguments but any permutation of $\ell_1$, $\ell_2$, $\ell_3$ and $\ell_4$
within the $(12)(34)$ pairing generates an equivalent trispectrum.
The flat-sky fully-reduced trispectrum
is related to its spherical equivalent by~\cite{HuAng}
\begin{equation}
\fullred^{\ell_1 \ell_2}_{\ell_3 \ell_4}(L) = \frac{1}{4\pi}
\twolpomul{L}^2\twolpomul{\ell_1\ell_2 \ell_3 \ell_4}
\ThreeJ{\ell_1}{\ell_2}{L}{0}{0}{0}
\ThreeJ{\ell_3}{\ell_4}{L}{0}{0}{0} \fullred^{(\ell_1 \ell_2)}_{(\ell_3 \ell_4)}(L) .
\label{eq:appb3}
\end{equation}

At second order in $\phi$, the flat-sky trispectrum is
\begin{equation}
T(\elb_1,\elb_2,\elb_3,\elb_4) = \frac{1}{2}
C_{\ell_2}^{\ThetaU \ThetaU} C_{\ell_4}^{\ThetaU \ThetaU} C_{|\elb_1 + \elb_2|}^{\phi\phi} (\elb_1+\elb_2)\cdot \elb_2 \, (\elb_3+\elb_4) \cdot \elb_4 + \text{perms.}
\label{eq:appb4}
\end{equation}
Breaking up the delta-function in Eq.~(\ref{eq:appb1}), we have
\begin{equation}
\langle \T(\elb_1) \T(\elb_2) \T(\elb_3) \T(\elb_4) \rangle_{C} =
\frac{1}{2} \int \frac{d^2 \vL}{(2\pi)^2} \delta(\elb_1 + \elb_2 + \vL)
\delta(\elb_3 + \elb_4 - \vL) C_{\ell_2}^{\ThetaU \ThetaU} C_{\ell_4}^{\ThetaU \ThetaU}C_{L}^{\phi\phi} (-\vL \cdot \elb_2) \vL \cdot \elb_4 + \text{perms.} , 
\label{eq:appb5}
\end{equation}
from which we can read off the fully-reduced trispectrum:
\begin{equation}
\fullred^{(\ell_1 \ell_2)}_{(\ell_3 \ell_4)}(L) = \frac{1}{4} C_{L}^{\phi\phi}
C_{\ell_2}^{\ThetaU \ThetaU} C_{\ell_4}^{\ThetaU \ThetaU} (L^2-\ell_1^2 + \ell_2^2)
(L^2 - \ell_3^2 + \ell_4^2) .
\label{eq:appb6}
\end{equation}
Making use of Eq.~(\ref{eq:appb3}), in the limit of large arguments [so that
e.g.\ $\ell^2 \approx \ell(\ell+1)$], we recover the spherical result in
Eq.~(\ref{eqn:lenstrispect}).

At ${\cal O}(\phi^4)$, there are four types of term that contribute to the
trispectrum:
\begin{displaymath} 
\begin{array}{lll}
\stackrel{\delta \Theta \delta \Theta \delta \Theta \delta \Theta}{
T(\elb_1, \elb_2, \elb_3, \elb_4)} &=& 
\frac{1}{4}\int\frac{d^2 \elb}{(2\pi)^2} \Bigl[\elb \cdot (\elb_1 - \elb)\ 
\elb \cdot (\elb_2 + \elb)
(\elb_1 + \elb_3 - \elb) \cdot (\elb_1 - \elb) 
(\elb_1 + \elb_3 - \elb) \cdot (\elb_2 + \elb) \\
& & \mbox{} \times 
C^{\phi\phi}_{|\elb - \elb_1|} 
C^{\phi\phi}_{|\elb + \elb_2|}
C^{\tilde{\Theta} \tilde{\Theta}}_{\ell}
C^{\tilde{\Theta} \tilde{\Theta}}_{|\elb_1 + \elb_3 - \elb|} \Bigr] + \text{perms.}\\
\stackrel{\delta^2 \Theta \delta^2 \Theta \tilde{\Theta} \tilde{\Theta}}{
T(\elb_1, \elb_2, \elb_3, \elb_4)} &=& 
\frac{1}{4}
C_{\ell_3}^{\tilde{\Theta} \tilde{\Theta}} 
C_{\ell_4}^{\tilde{\Theta} \tilde{\Theta}} 
\int \frac{d^2 \elb}{(2\pi)^2} 
\elb_3 \cdot \elb \  \elb_4 \cdot \elb\ 
\elb_3 \cdot (\elb - \elb_1 -\elb_3) 
\elb_4 \cdot (\elb -\elb_1 -\elb_3) 
C^{\phi\phi}_{\ell} 
C^{\phi\phi}_{|\elb - \elb_1 -\elb_3|} + \text{perms.}\\
\stackrel{\delta^2 \Theta \delta \Theta \delta \Theta \tilde{\Theta}}{
T(\elb_1, \elb_2, \elb_3, \elb_4) }
&=& - \frac{1}{2} C^{\phi\phi}_{|\elb_2+\elb_4|} R\, \ell_1^2 
       C_{\ell_1}^{\tilde{\Theta} \tilde{\Theta}}\, \elb_1 \cdot (\elb_1 + \elb_3) 
       C_{\ell_4}^{\tilde{\Theta} \tilde{\Theta}}\, \elb_4 \cdot (\elb_2 + \elb_4)  \\
 & & \mbox{} + C^{\phi\phi}_{|\elb_2 + \elb_4|} 
       C_{\ell_4}^{\tilde{\Theta} \tilde{\Theta}} \, 
       \elb_4 \cdot (\elb_2 + \elb_4) 
       \int \frac{d^2 \elb}{(2\pi)^2} 
       \elb \cdot (\elb_1 +\elb_3) 
       \left[ \elb\cdot(\elb -\elb_3) \right]^2 
       C^{\tilde{\Theta} \tilde{\Theta}}_{\ell}
       C^{\phi\phi}_{|\elb - \elb_3|}  \\ 
 & & \mbox{} - \frac{1}{2} C_{\ell_1}^{\tilde{\Theta} \tilde{\Theta}}
\int \frac{d^2 \elb}{(2\pi)^2} \elb_1 \cdot (\elb + \elb_3) \elb\cdot
(\elb+ \elb_3) \elb_1 \cdot (\elb-\elb_2) \elb\cdot(\elb-\elb_2)
C^{\tilde{\Theta} \tilde{\Theta}}_{\ell}
C^{\phi\phi}_{|\elb-\elb_2|}C^{\phi\phi}_{|\elb +\elb_3|} + \text{perms.}
\\
\stackrel{\delta^3 \Theta \delta \Theta \tilde{\Theta} \tilde{\Theta}}{
T(\elb_1, \elb_2, \elb_3, \elb_4) } &=& 
- \frac{1}{2} C^{\phi\phi}_{|\elb_2+\elb_4|} R\, \ell_1^2 
       C_{\ell_1}^{\tilde{\Theta} \tilde{\Theta}}\, \elb_1 \cdot (\elb_1 + \elb_3) 
       C_{\ell_4}^{\tilde{\Theta} \tilde{\Theta}}\, \elb_4 \cdot (\elb_2 + \elb_4)
+ \text{perms.} ,
\end{array}
\label{eqn:4thordertermsflat}
\end{displaymath}
where 
\begin{equation}
R \equiv \frac{1}{4\pi} \int L^3 C^{\phi\phi}_L dL 
\end{equation}
is half the variance of the deflection field.

The subset of trispectrum terms that dominate the power spectrum of the
reconstructed lensing field are
\begin{eqnarray}
\overset{\rm{dom}}{T}(\elb_1,\elb_2,\elb_3,\elb_4) &=&
-C^{\phi\phi}_{|\elb_3+\elb_4|} R\, \ell_1^2 
       C_{\ell_1}^{\tilde{\Theta} \tilde{\Theta}}\, \elb_1 \cdot (\elb_1 + \elb_2) 
       C_{\ell_3}^{\tilde{\Theta} \tilde{\Theta}}\, \elb_3 \cdot (\elb_3 + \elb_4)
\nonumber \\
&& \mbox{} + C^{\phi\phi}_{|\elb_3 + \elb_4|} 
       C_{\ell_3}^{\tilde{\Theta} \tilde{\Theta}} \, 
       \elb_3 \cdot (\elb_3 + \elb_4) 
       \int \frac{d^2 \elb}{(2\pi)^2} 
       \elb \cdot (\elb_1 +\elb_2) 
       \left[ \elb\cdot(\elb -\elb_1) \right]^2 
       C^{\tilde{\Theta} \tilde{\Theta}}_{\ell}
       C^{\phi\phi}_{|\elb - \elb_1|} \nonumber \\
&& \mbox{} + \text{perms.} ,
\label{eq:flatdom}
\end{eqnarray}
where we have chosen this particular permutation to display for later
convenience. The first term (plus its permutations) generates a fully-reduced
trispectrum
\begin{equation}
\fullred^{(\ell_1 \ell_2)}_{(\ell_3 \ell_4)}(L) = - \frac{1}{2}
C_L^{\phi\phi} R \ell_1^2 C_{\ell_1}^{\ThetaU \ThetaU}
(L^2+\ell_1^2-\ell_2^2)C_{\ell_3}^{\ThetaU \ThetaU}
(L^2+\ell_3^2-\ell_4^2) .
\label{eq:flatred2}
\end{equation}
Using Eq.~(\ref{eq:appb3}), this gives a full-sky trispectrum which
for large multipoles reduces simply to the second term in Eq.~(\ref{eqn:domofterms}). The second term on the right of Eq.~(\ref{eq:flatdom}) gives a fully-reduced reduced trispectrum
\begin{eqnarray}
\fullred^{(\ell_1 \ell_2)}_{(\ell_3 \ell_4)}(L) &=& - 2
C_L^{\phi\phi} C_{\ell_3}^{\ThetaU \ThetaU} \elb_3 \cdot \vL \int
\frac{d^2 \elb_a}{(2\pi)^2} \elb_a \cdot \vL [\elb_a \cdot (\elb_a - \elb_1)]^2
C_{\ell_a}^{\ThetaU \ThetaU} C^{\phi\phi}_{|\elb_a - \elb_1|} \nonumber \\
&=&  - 2 C_L^{\phi\phi} C_{\ell_3}^{\ThetaU \ThetaU} \elb_3 \cdot \vL
\elb_1 \cdot \vL \int \frac{d^2 \elb_a}{(2\pi)^2}
\frac{\elb_a \cdot \elb_1}{\el_1^2} [\elb_a \cdot (\elb_a - \elb_1)]^2
C_{\ell_a}^{\ThetaU \ThetaU} C^{\phi\phi}_{|\elb_a - \elb_1|},
\label{eq:flatred1}
\end{eqnarray}
where $\elb_3 + \elb_4 = \vL = -(\elb_1 + \elb_2)$.
To write this in a form that manifestly depends only on $\ell_1$, $\ell_2$,
$\ell_3$, $\ell_4$ and $L$ for comparison to the full-sky result, we
write $C^{\phi\phi}_{|\elb_a - \elb_1|} = \int d^2 \elb_b \delta(\elb_a + \elb_b
- \elb_1) C_{\ell_b}^{\phi\phi}$ and use the large-$\ell$ expansion of the
delta function (e.g.~\cite{HuAng}),
\begin{equation}
\delta(\elb_a + \elb_b - \elb_1) \approx \frac{1}{\pi} \sum_{m_a m_b m_1}
(-1)^{m_1} e^{i m_a \phi_{\elb_a}} e^{i m_b \phi_{\elb_b}} e^{-i m_1 \phi_{\elb_1}}
\ThreeJ{\ell_a}{\ell_b}{\ell_1}{0}{0}{0}
\ThreeJ{\ell_a}{\ell_b}{\ell_1}{m_a}{m_b}{-m_1} ,
\end{equation}
where, for example, $\phi_{\elb_1}$ is the angle that $\elb_1$ makes with the
$x$-axis. Substituting into Eq.~(\ref{eq:flatred1}), integrating over
$\phi_{\elb_a}$ and $\phi_{\elb_b}$ and simplifying with recursion relations
for the $3j$ symbols, we find
\begin{equation}
\fullred^{(\ell_1 \ell_2)}_{(\ell_3 \ell_4)}(L) \approx  \frac{1}{4}
C_L^{\phi\phi} C_{\ell_3}^{\ThetaU \ThetaU} (\ell_4^2 - \ell_3^2 - L^2)
(\el_2^2-\ell_1^2-L^2) \sum_{\ell_a \ell_b} C_{\ell_a}^{\ThetaU \ThetaU} C^{\phi\phi}_{\ell_b} \frac{(\ell_1^2 + \ell_a^2 - \ell_b^2)}{\ell_1^2
\twolpomul{\ell_1}^2} (F_{\ell_1 \ell_a \ell_b})^2 .
\end{equation}
This generates a full-sky trispectrum which for large multipoles
reduces to the first term in Eq.~(\ref{eqn:domofterms}).

For lens reconstruction, the standard optimal estimator is given on the flat-sky by~\cite{HuMap}
\begin{equation}
\hat{\phi}(\mathbf{L}) = {\cal A}_{L} \int\frac{d^2 \elb_1}{2\pi} \ThetaL(\elb_1) \ThetaL(\elb_2) g_{\elb_1 \elb_2}(\mathbf{L}) ,
\end{equation}
where $\mathbf{L} = \elb_1 + \elb_2$ and
\begin{eqnarray}
g_{\elb_1 \elb_2}(\mathbf{L}) &=& \frac{f_{\elb_1 \elb_2} (\mathbf{L})}{2 C_{\el_1, {\rm expt}}^{\Theta \Theta} 
C_{\el_2, {\rm expt}}^{\Theta \Theta} } \ , \\
f_{\elb_1 \elb_2} (\mathbf{L}) &=&  \elb_1 \cdot \mathbf{L} C_{\el_1}^{\tilde{\Theta} \tilde{\Theta}} + 
\elb_2 \cdot \mathbf{L} C_{\el_2}^{\tilde{\Theta} \tilde{\Theta}}\ .
\end{eqnarray}
The normalization is
\begin{equation}
\mathcal{A}_L^{-1} = \int \frac{d^2 \elb_1}{(2\pi)^2}
f_{\elb_1 \elb_2}(\vL) g_{\elb_1\elb_2}(\vL) .
\end{equation}
The power spectrum of the reconstruction involves the four-point function
of the observed CMB; it can be written in terms of the fully-reduced
trispectrum as
\begin{equation}
\langle C_L^{\hat{\phi}\hat{\phi}}\rangle = N_L^{(0)} + 4 \mathcal{A}_L^2
\int \frac{d^2 \elb_1}{(2\pi)^2} \frac{d^2 \elb_3}{(2\pi)^2}
g_{\elb_1 \elb_2}(\vL) g_{\elb_3 \elb_4}(\vL) \left[
\fullred^{(\ell_1 \ell_2)}_{(\ell_3 \ell_4)}(L) +
\fullred^{(\ell_1 \ell_3)}_{(\ell_2 \ell_4)}(|\elb_1-\elb_3|) + 
\fullred^{(\ell_1 \ell_4)}_{(\ell_2 \ell_3)}(|\elb_1-\elb_4|) \right] ,
\label{eq:flatpower}
\end{equation}
where $\elb_1 + \elb_2 = \vL = \elb_3 + \elb_4$.
Here, $N_L^{(0)} = \mathcal{A}_L$ arises from the Gaussian (i.e.\ disconnected)
part of the four-point function. To second order in $\phi$, the
primary coupling term in Eq.~(\ref{eq:flatpower}) gives the lensing power
spectrum, $C_\ell^{\phi\phi}$ that we aim to reconstruct. The other two couplings
give the $N^{(1)}_L$ bias found in Ref.~\cite{KesCooKam}:
\begin{eqnarray}
N^{(1)}_L &=& -4 \mathcal{A}_L^2 \int \frac{d^2 \elb_1}{(2\pi)^2}
\frac{d^2 \elb_3}{(2\pi)^2} g_{\elb_1 \elb_2}(\vL) g_{\elb_3 \elb_4}(\vL)
\Bigl[ \elb_1 \cdot(\elb_1-\elb_3)\elb_2 \cdot(\elb_1-\elb_3) C^{\phi\phi}_{|\elb_1-\elb_3|} C^{\tilde{\Theta}\tilde{\Theta}}_{\ell_1}
C^{\tilde{\Theta}\tilde{\Theta}}_{\ell_2} \nonumber \\
&&\mbox{} \hspace{0.3\textwidth}
+ \elb_1 \cdot(\elb_1-\elb_4)(-\elb_3) \cdot(\elb_1-\elb_4) C^{\phi\phi}_{|\elb_1-\elb_4|} C^{\tilde{\Theta}\tilde{\Theta}}_{\ell_1}
C^{\tilde{\Theta}\tilde{\Theta}}_{\ell_3} \Bigr] .
\end{eqnarray}
To fourth order in $\phi$, the primary coupling of the dominant trispectrum
terms, Eqs.~(\ref{eq:flatred2}) and~(\ref{eq:flatred1}), give the $N_L^{(2)}$
bias:
\begin{equation}
N_{L}^{(2)} \approx  4 C^{\phi\phi}_L \mathcal{A}_{L}^2 
\left( \int \frac{d^2 \elb_3}{(2\pi)^2} g_{\elb_3 \elb_4}(\vL)
f_{\elb_3 \elb_4}(\vL)\right)
\int \frac{d^2 \elb_1}{(2\pi)^2}
g_{\elb_1 \elb_2}(\vL) \Biggl(
\int \frac{d^2 \elb}{(2\pi)^2} \elb\cdot \vL [\elb\cdot(\elb-\elb_1)]^2 C_\ell^{\tilde{\Theta}\tilde{\Theta}}
C_{|\elb-\elb_1|}^{\phi\phi}
- R \elb_1 \cdot \vL \ell_1^2 C_{\ell_1}^{\tilde{\Theta}\tilde{\Theta}} 
\Biggr).
\label{eq:flatN2}
\end{equation}
This is the flat-sky version of Eq.~(\ref{eqn:nl2bias}).
We show in the text that $N^{(2)}_L/C_L^{\phi\phi}$ can be expressed in terms of the leading-order difference between the lensed and unlensed power spectra.
More correctly,
the final term in Eq.~(\ref{eq:flatN2}) can be shown to involve the difference
between the power spectrum of the unlensed CMB and the 
cross spectrum of the lensed CMB with the lensed temperature gradient,
but the latter is equal to the lensed spectrum to better than one
percent (Lewis, Challinor \& Hanson, in prep.).
We show further in the text that $N_L^{(2)}/2$
also arises when computing the power spectrum $C_L^{\phi\hat{\phi}}$.
Using techniques from Lewis, Challinor \& Hanson (in prep.),
the latter
can be computed non-perturbatively in the lensing deflection as
\begin{equation}
\langle C_L^{\phi\hat{\phi}} \rangle \approx \mathcal{A}_L C_L^{\phi\phi} \int \frac{d^2 \elb_1}{(2\pi)^2}g_{\elb_1 \elb_2}(\vL) \left(\elb_1\cdot \vL
C_{\el_1}^{\Theta \Theta} + \elb_2 \cdot \vL
C_{\el_2}^{\Theta \Theta} \right) ,
\end{equation}
where the approximation arises from replacing the exact cross power spectrum
of the lensed temperature and the lensed temperature gradient with the exact
$C_{\el}^{\Theta \Theta}$.
We expect $\langle C_L^{\phi\hat{\phi}} \rangle$
to equal $C_L^{\phi\phi}$ plus half the non-perturbative
generalization of $N_L^{(2)}$. We thus expect the dominant low-$\ell$
fractional bias in $C_L^{\hat{\phi}\hat{\phi}}$
to be controlled at higher order by the difference between the (exact) lensed
spectrum and the unlensed spectrum.

\bibliography{quad_phest_biases_bib_manual}
\label{lastpage}
\end{document}

%% file: macros.tex
\renewcommand{\,}{\thinspace}

\newcommand{\vL}{\mathbf{L}}

\newcommand{\vc}[1]{\ensuremath{\bm{#1}}}

\newcommand{\Ylm}[1]{Y_{\el_#1}^{\m_#1}}
\newcommand{\spin}[1]{\,{}_{#1}^{\vphantom{m}}}
\newcommand{\Ylmn}{Y_{\el}^{\m}}
\newcommand{\Yslmn}{\spin{s} Y_{\el}^{m}}
\newcommand{\Yslm}[3]{\spin{#1} Y_{#2}^{#3}}

\newcommand{\trispect}{{\cal T}}

\newcommand{\T}{\Theta}
\newcommand{\ThreeJ}[6]{\left(
                           \begin{array}{ccc}
        \! #1\! & #2\!  & #3\!  \\
        \! #4\! & #5\!  & #6\!
                           \end{array}
                   \right)}

\newcommand{\wsixj}[6]{\left\{
                           \begin{array}{ccc}
         #1 & #2  & #3  \\
         #4 & #5  & #6
                           \end{array}
                   \right\}}

\newcommand{\WMAP}{{\sl WMAP\/}\ }
\newcommand{\Planck}{{\sl Planck\/}\ }

\newcommand{\healpix}{{\sc HEALPix}\ }
\newcommand{\camb}{\textsc{CAMB}}

\newcommand{\el}{\ell}
\newcommand{\elb}{\mathbf{l}}

\newcommand{\es}{s}
\newcommand{\m}{m}

\newcommand{\intsky}[1]{\int d\Omega #1}

\newcommand{\itn}{I_{\el \el_1 \el_2}^{\m \m_1 \m_2}}
\newcommand{\jtn}{J_{\el \el_1 \el_2 \el_3}^{\m \m_1 \m_2 \m_3}}
\newcommand{\ktn}{K_{\el \el_1 \el_2 \el_3 \el_4}^{\m \m_1 \m_2 \m_3 \m_4}}

\newcommand{\gYlm}[2]{\nabla_{#1}\Ylm{#2}}

\newcommand{\etheta}{\hat{\vc{e}}_{\theta}}
\newcommand{\ephi}{\hat{\vc{e}}_{\phi}}
\newcommand{\epm}{\vc{e}_{\pm}}

\newcommand{\lbar}[1]{\underline{\el}_{#1}}

\newcommand{\fourpt}{\langle\T_{\lbar{1}}\T_{\lbar{2}}\T_{\lbar{3}}\T_{\lbar{4}} \rangle}
\newcommand{\fourpta}[4]{\langle\T_{\lbar{#1}}\T_{\lbar{#2}}\T_{\lbar{#3}}\T_{\lbar{#4}} \rangle}
\newcommand{\eightpt}{\langle\T_{\lbar{1}}\T_{\lbar{2}}\T_{\lbar{3}}\T_{\lbar{4}}\T_{\lbar{5}}\T_{\lbar{6}}\T_{\lbar{7}}\T_{\lbar{8}} \rangle}
\newcommand{\prim}[1]{#1^{\el_1 \el_2}_{\el_3 \el_4}}
\newcommand{\sfourls}{\sum_{\lbar{1}\;\lbar{2}\;\lbar{3}\;\lbar{4}}}
\newcommand{\sfourlsnobar}{\sum_{\el_{1}\;\el_{2}\;\el_{3}\;\el_{4}}}
\newcommand{\stwols}{\sum_{\lbar{1}\;\lbar{2}}}

\newcommand{\cppest}[1]{C^{\phiest{}\phiest{}}_{#1}}
\newcommand{\phiest}[1]{\hat{\phi_{#1}}}

\newcommand{\gwt}[2]{g_{\el_{#1}\el_{#2}}}

\newcommand{\fwt}[2]{f_{\el_{#1}L\el_{#2}}}
\newcommand{\lft}[3]{f_{{#1}{#2}{#3}}}
\newcommand{\bfftL}[2]{{{F}}_{{\el_{#1}}L{\el_{#2}}}}
\newcommand{\bfft}[3]{{{F}}_{{\el_{#1}}{\el_{#2}}{\el_{#3}}}}
\newcommand{\bfftp}[3]{{{F}}_{{{#1}}{{#2}}{{#3}}}}

\newcommand{\cttlp}[1]{C^{\Theta \Theta}_{#1}}

\newcommand{\cttup}[1]{C^{\tilde{\Theta} \tilde{\Theta}}_{#1}}
\newcommand{\cttu}[1]{C^{\tilde{\Theta} \tilde{\Theta}}_{\el_{#1}}}

\newcommand{\clexptp}[1]{\cttlp{#1,\rm{expt}}}

\newcommand{\ThetaL}{\Theta}
\newcommand{\ThetaU}{\tilde{\Theta}}

\newcommand{\cpp}[1]{C^{\phi\phi}_{\el_{#1}}}
\newcommand{\cppp}[1]{C^{\phi\phi}_{{#1}}}

\newcommand{\twolpomul}[1]{\Uppi_{#1}}
\newcommand{\llponemul}[1]{\Upxi_{#1}}

\newcommand{\telm}{\Theta_{\el \m}}
\newcommand{\telmn}[1]{\Theta_{\lbar{#1}}}

\newcommand{\ttelmn}[1]{\tilde{\Theta}_{\lbar{#1}}}

\newcommand{\Yslmp}[2]{\spin{#1} Y_{\el_{#2}}^{\m_#2}}

\newcommand{\fullred}{{\mathbb{T}}}

\newcommand{\cov}{{\rm Cov}}
\newcommand{\var}{{\rm Var}}
\newcommand{\rcov}{\ensuremath{\mathcal{R}}}
\newcommand{\order}{\ensuremath{\mathcal{O}}}

\newcommand{\jhookn}{\ominus}
\newcommand{\jfishn}{\otimes}
\newcommand{\jhook}[5]{\jhookn^{{#1}{#2}}_{{#3}{#4}} (#5)}
\newcommand{\jfish}[5]{\jfishn^{{#1}{#2}}_{{#3}{#4}} (#5)}

\newcommand{\fwhm}{\ensuremath{\sigma_{\rm FWHM}}}

%% file: quad_phest_biases.bbl
\begin{thebibliography}{37}
\expandafter\ifx\csname natexlab\endcsname\relax\def\natexlab#1{#1}\fi
\expandafter\ifx\csname bibnamefont\endcsname\relax
  \def\bibnamefont#1{#1}\fi
\expandafter\ifx\csname bibfnamefont\endcsname\relax
  \def\bibfnamefont#1{#1}\fi
\expandafter\ifx\csname citenamefont\endcsname\relax
  \def\citenamefont#1{#1}\fi
\expandafter\ifx\csname url\endcsname\relax
  \def\url#1{\texttt{#1}}\fi
\expandafter\ifx\csname urlprefix\endcsname\relax\def\urlprefix{URL }\fi
\providecommand{\bibinfo}[2]{#2}
\providecommand{\eprint}[2][]{\url{#2}}

\bibitem[{\citenamefont{{Komatsu} et~al.}(2010)\citenamefont{{Komatsu},
  {Smith}, {Dunkley}, {Bennett}, {Gold}, {Hinshaw}, {Jarosik}, {Larson},
  {Nolta}, {Page} et~al.}}]{2010arXiv1001.4538K}
\bibinfo{author}{\bibfnamefont{E.}~\bibnamefont{{Komatsu}}},
  \bibinfo{author}{\bibfnamefont{K.~M.} \bibnamefont{{Smith}}},
  \bibinfo{author}{\bibfnamefont{J.}~\bibnamefont{{Dunkley}}},
  \bibinfo{author}{\bibfnamefont{C.~L.} \bibnamefont{{Bennett}}},
  \bibinfo{author}{\bibfnamefont{B.}~\bibnamefont{{Gold}}},
  \bibinfo{author}{\bibfnamefont{G.}~\bibnamefont{{Hinshaw}}},
  \bibinfo{author}{\bibfnamefont{N.}~\bibnamefont{{Jarosik}}},
  \bibinfo{author}{\bibfnamefont{D.}~\bibnamefont{{Larson}}},
  \bibinfo{author}{\bibfnamefont{M.~R.} \bibnamefont{{Nolta}}},
  \bibinfo{author}{\bibfnamefont{L.}~\bibnamefont{{Page}}},
  \bibnamefont{et~al.}, \bibinfo{journal}{ArXiv e-prints}
  (\bibinfo{year}{2010}), \eprint{1001.4538}.

\bibitem[{\citenamefont{{Bartolo} et~al.}(2004)\citenamefont{{Bartolo},
  {Komatsu}, {Matarrese}, and {Riotto}}}]{2004PhR...402..103B}
\bibinfo{author}{\bibfnamefont{N.}~\bibnamefont{{Bartolo}}},
  \bibinfo{author}{\bibfnamefont{E.}~\bibnamefont{{Komatsu}}},
  \bibinfo{author}{\bibfnamefont{S.}~\bibnamefont{{Matarrese}}},
  \bibnamefont{and} \bibinfo{author}{\bibfnamefont{A.}~\bibnamefont{{Riotto}}},
  \bibinfo{journal}{\physrep} \textbf{\bibinfo{volume}{402}},
  \bibinfo{pages}{103} (\bibinfo{year}{2004}), \eprint{arXiv:astro-ph/0406398}.

\bibitem[{\citenamefont{Lewis and Challinor}(2006)}]{LewChaLens}
\bibinfo{author}{\bibfnamefont{A.}~\bibnamefont{Lewis}} \bibnamefont{and}
  \bibinfo{author}{\bibfnamefont{A.}~\bibnamefont{Challinor}},
  \bibinfo{journal}{Physics Reports} \textbf{\bibinfo{volume}{429}},
  \bibinfo{pages}{1} (\bibinfo{year}{2006}),
  \urlprefix\url{doi:10.1016/j.physrep.2006.03.002}.

\bibitem[{\citenamefont{{Seljak}}(1996)}]{1996ApJ...463....1S}
\bibinfo{author}{\bibfnamefont{U.}~\bibnamefont{{Seljak}}},
  \bibinfo{journal}{\apj} \textbf{\bibinfo{volume}{463}}, \bibinfo{pages}{1}
  (\bibinfo{year}{1996}), \eprint{arXiv:astro-ph/9505109}.

\bibitem[{\citenamefont{{Zaldarriaga}}(2000)}]{2000PhRvD..62f3510Z}
\bibinfo{author}{\bibfnamefont{M.}~\bibnamefont{{Zaldarriaga}}},
  \bibinfo{journal}{\prd} \textbf{\bibinfo{volume}{62}},
  \bibinfo{pages}{063510} (\bibinfo{year}{2000}),
  \eprint{arXiv:astro-ph/9910498}.

\bibitem[{\citenamefont{{Goldberg} and {Spergel}}(1999)}]{1999PhRvD..59j3002G}
\bibinfo{author}{\bibfnamefont{D.~M.} \bibnamefont{{Goldberg}}}
  \bibnamefont{and} \bibinfo{author}{\bibfnamefont{D.~N.}
  \bibnamefont{{Spergel}}}, \bibinfo{journal}{\prd}
  \textbf{\bibinfo{volume}{59}}, \bibinfo{pages}{103002}
  (\bibinfo{year}{1999}), \eprint{arXiv:astro-ph/9811251}.

\bibitem[{\citenamefont{{Hanson} et~al.}(2009)\citenamefont{{Hanson}, {Smith},
  {Challinor}, and {Liguori}}}]{2009PhRvD..80h3004H}
\bibinfo{author}{\bibfnamefont{D.}~\bibnamefont{{Hanson}}},
  \bibinfo{author}{\bibfnamefont{K.~M.} \bibnamefont{{Smith}}},
  \bibinfo{author}{\bibfnamefont{A.}~\bibnamefont{{Challinor}}},
  \bibnamefont{and}
  \bibinfo{author}{\bibfnamefont{M.}~\bibnamefont{{Liguori}}},
  \bibinfo{journal}{\prd} \textbf{\bibinfo{volume}{80}},
  \bibinfo{pages}{083004} (\bibinfo{year}{2009}), \eprint{0905.4732}.

\bibitem[{\citenamefont{{Hu}}(2002)}]{2002PhRvD..65b3003H}
\bibinfo{author}{\bibfnamefont{W.}~\bibnamefont{{Hu}}}, \bibinfo{journal}{\prd}
  \textbf{\bibinfo{volume}{65}}, \bibinfo{pages}{023003}
  (\bibinfo{year}{2002}), \eprint{arXiv:astro-ph/0108090}.

\bibitem[{\citenamefont{{Smith} et~al.}(2008)\citenamefont{{Smith}, {Cooray},
  {Das}, {Dor{\'e}}, {Hanson}, {Hirata}, {Kaplinghat}, {Keating}, {LoVerde},
  {Miller} et~al.}}]{2008arXiv0811.3916S}
\bibinfo{author}{\bibfnamefont{K.~M.} \bibnamefont{{Smith}}},
  \bibinfo{author}{\bibfnamefont{A.}~\bibnamefont{{Cooray}}},
  \bibinfo{author}{\bibfnamefont{S.}~\bibnamefont{{Das}}},
  \bibinfo{author}{\bibfnamefont{O.}~\bibnamefont{{Dor{\'e}}}},
  \bibinfo{author}{\bibfnamefont{D.}~\bibnamefont{{Hanson}}},
  \bibinfo{author}{\bibfnamefont{C.}~\bibnamefont{{Hirata}}},
  \bibinfo{author}{\bibfnamefont{M.}~\bibnamefont{{Kaplinghat}}},
  \bibinfo{author}{\bibfnamefont{B.}~\bibnamefont{{Keating}}},
  \bibinfo{author}{\bibfnamefont{M.}~\bibnamefont{{LoVerde}}},
  \bibinfo{author}{\bibfnamefont{N.}~\bibnamefont{{Miller}}},
  \bibnamefont{et~al.}, \bibinfo{journal}{ArXiv e-prints}
  (\bibinfo{year}{2008}), \eprint{0811.3916}.

\bibitem[{\citenamefont{{Stompor} and
  {Efstathiou}}(1999)}]{1999MNRAS.302..735S}
\bibinfo{author}{\bibfnamefont{R.}~\bibnamefont{{Stompor}}} \bibnamefont{and}
  \bibinfo{author}{\bibfnamefont{G.}~\bibnamefont{{Efstathiou}}},
  \bibinfo{journal}{\mnras} \textbf{\bibinfo{volume}{302}},
  \bibinfo{pages}{735} (\bibinfo{year}{1999}), \eprint{arXiv:astro-ph/9805294}.

\bibitem[{\citenamefont{{Smith}
  et~al.}(2006{\natexlab{a}})\citenamefont{{Smith}, {Challinor}, and
  {Rocha}}}]{2006PhRvD..73b3517S}
\bibinfo{author}{\bibfnamefont{S.}~\bibnamefont{{Smith}}},
  \bibinfo{author}{\bibfnamefont{A.}~\bibnamefont{{Challinor}}},
  \bibnamefont{and} \bibinfo{author}{\bibfnamefont{G.}~\bibnamefont{{Rocha}}},
  \bibinfo{journal}{\prd} \textbf{\bibinfo{volume}{73}},
  \bibinfo{pages}{023517} (\bibinfo{year}{2006}{\natexlab{a}}),
  \eprint{arXiv:astro-ph/0511703}.

\bibitem[{\citenamefont{{Smith}
  et~al.}(2006{\natexlab{b}})\citenamefont{{Smith}, {Hu}, and
  {Kaplinghat}}}]{2006PhRvD..74l3002S}
\bibinfo{author}{\bibfnamefont{K.~M.} \bibnamefont{{Smith}}},
  \bibinfo{author}{\bibfnamefont{W.}~\bibnamefont{{Hu}}}, \bibnamefont{and}
  \bibinfo{author}{\bibfnamefont{M.}~\bibnamefont{{Kaplinghat}}},
  \bibinfo{journal}{\prd} \textbf{\bibinfo{volume}{74}},
  \bibinfo{pages}{123002} (\bibinfo{year}{2006}{\natexlab{b}}),
  \eprint{arXiv:astro-ph/0607315}.

\bibitem[{\citenamefont{{Lesgourgues} et~al.}(2006)\citenamefont{{Lesgourgues},
  {Perotto}, {Pastor}, and {Piat}}}]{LesPerPasPia}
\bibinfo{author}{\bibfnamefont{J.}~\bibnamefont{{Lesgourgues}}},
  \bibinfo{author}{\bibfnamefont{L.}~\bibnamefont{{Perotto}}},
  \bibinfo{author}{\bibfnamefont{S.}~\bibnamefont{{Pastor}}}, \bibnamefont{and}
  \bibinfo{author}{\bibfnamefont{M.}~\bibnamefont{{Piat}}},
  \bibinfo{journal}{\prd} \textbf{\bibinfo{volume}{73}},
  \bibinfo{pages}{045021} (\bibinfo{year}{2006}),
  \eprint{arXiv:astro-ph/0511735}.

\bibitem[{\citenamefont{{Challinor} and {Chon}}(2002)}]{2002PhRvD..66l7301C}
\bibinfo{author}{\bibfnamefont{A.}~\bibnamefont{{Challinor}}} \bibnamefont{and}
  \bibinfo{author}{\bibfnamefont{G.}~\bibnamefont{{Chon}}},
  \bibinfo{journal}{\prd} \textbf{\bibinfo{volume}{66}},
  \bibinfo{pages}{127301} (\bibinfo{year}{2002}),
  \eprint{arXiv:astro-ph/0301064}.

\bibitem[{\citenamefont{{Challinor} and {Lewis}}(2005)}]{ChaLewCorr}
\bibinfo{author}{\bibfnamefont{A.}~\bibnamefont{{Challinor}}} \bibnamefont{and}
  \bibinfo{author}{\bibfnamefont{A.}~\bibnamefont{{Lewis}}},
  \bibinfo{journal}{\prd} \textbf{\bibinfo{volume}{71}},
  \bibinfo{pages}{103010} (\bibinfo{year}{2005}),
  \eprint{arXiv:astro-ph/0502425}.

\bibitem[{\citenamefont{{Okamoto} and {Hu}}(2003)}]{OkHu}
\bibinfo{author}{\bibfnamefont{T.}~\bibnamefont{{Okamoto}}} \bibnamefont{and}
  \bibinfo{author}{\bibfnamefont{W.}~\bibnamefont{{Hu}}},
  \bibinfo{journal}{\prd} \textbf{\bibinfo{volume}{67}},
  \bibinfo{pages}{083002} (\bibinfo{year}{2003}),
  \eprint{arXiv:astro-ph/0301031}.

\bibitem[{\citenamefont{{Smith} et~al.}(2007)\citenamefont{{Smith}, {Zahn}, and
  {Dor{\'e}}}}]{SmiZahDor}
\bibinfo{author}{\bibfnamefont{K.~M.} \bibnamefont{{Smith}}},
  \bibinfo{author}{\bibfnamefont{O.}~\bibnamefont{{Zahn}}}, \bibnamefont{and}
  \bibinfo{author}{\bibfnamefont{O.}~\bibnamefont{{Dor{\'e}}}},
  \bibinfo{journal}{\prd} \textbf{\bibinfo{volume}{76}},
  \bibinfo{pages}{043510} (\bibinfo{year}{2007}), \eprint{arXiv:0705.3980}.

\bibitem[{\citenamefont{{Hirata} et~al.}(2008)\citenamefont{{Hirata}, {Ho},
  {Padmanabhan}, {Seljak}, and {Bahcall}}}]{HirHoSel}
\bibinfo{author}{\bibfnamefont{C.~M.} \bibnamefont{{Hirata}}},
  \bibinfo{author}{\bibfnamefont{S.}~\bibnamefont{{Ho}}},
  \bibinfo{author}{\bibfnamefont{N.}~\bibnamefont{{Padmanabhan}}},
  \bibinfo{author}{\bibfnamefont{U.}~\bibnamefont{{Seljak}}}, \bibnamefont{and}
  \bibinfo{author}{\bibfnamefont{N.~A.} \bibnamefont{{Bahcall}}},
  \bibinfo{journal}{\prd} \textbf{\bibinfo{volume}{78}},
  \bibinfo{pages}{043520} (\bibinfo{year}{2008}), \eprint{0801.0644}.

\bibitem[{\citenamefont{{Kaplinghat} et~al.}(2003)\citenamefont{{Kaplinghat},
  {Knox}, and {Song}}}]{KapKnoSon}
\bibinfo{author}{\bibfnamefont{M.}~\bibnamefont{{Kaplinghat}}},
  \bibinfo{author}{\bibfnamefont{L.}~\bibnamefont{{Knox}}}, \bibnamefont{and}
  \bibinfo{author}{\bibfnamefont{Y.-S.} \bibnamefont{{Song}}},
  \bibinfo{journal}{Physical Review Letters} \textbf{\bibinfo{volume}{91}},
  \bibinfo{pages}{241301} (\bibinfo{year}{2003}),
  \eprint{arXiv:astro-ph/0303344}.

\bibitem[{\citenamefont{{Kesden} et~al.}(2003)\citenamefont{{Kesden}, {Cooray},
  and {Kamionkowski}}}]{KesCooKam}
\bibinfo{author}{\bibfnamefont{M.}~\bibnamefont{{Kesden}}},
  \bibinfo{author}{\bibfnamefont{A.}~\bibnamefont{{Cooray}}}, \bibnamefont{and}
  \bibinfo{author}{\bibfnamefont{M.}~\bibnamefont{{Kamionkowski}}},
  \bibinfo{journal}{\prd} \textbf{\bibinfo{volume}{67}},
  \bibinfo{pages}{123507} (\bibinfo{year}{2003}),
  \eprint{arXiv:astro-ph/0302536}.

\bibitem[{\citenamefont{{Amblard} et~al.}(2004)\citenamefont{{Amblard}, {Vale},
  and {White}}}]{AmbValWhi}
\bibinfo{author}{\bibfnamefont{A.}~\bibnamefont{{Amblard}}},
  \bibinfo{author}{\bibfnamefont{C.}~\bibnamefont{{Vale}}}, \bibnamefont{and}
  \bibinfo{author}{\bibfnamefont{M.}~\bibnamefont{{White}}},
  \bibinfo{journal}{New Astronomy} \textbf{\bibinfo{volume}{9}},
  \bibinfo{pages}{687} (\bibinfo{year}{2004}), \eprint{arXiv:astro-ph/0403075}.

\bibitem[{\citenamefont{{Lewis}}(2005)}]{LewLpix}
\bibinfo{author}{\bibfnamefont{A.}~\bibnamefont{{Lewis}}},
  \bibinfo{journal}{\prd} \textbf{\bibinfo{volume}{71}},
  \bibinfo{pages}{083008} (\bibinfo{year}{2005}),
  \eprint{arXiv:astro-ph/0502469}.

\bibitem[{\citenamefont{{Knox}}(1995)}]{KnoxDet}
\bibinfo{author}{\bibfnamefont{L.}~\bibnamefont{{Knox}}},
  \bibinfo{journal}{\prd} \textbf{\bibinfo{volume}{52}}, \bibinfo{pages}{4307}
  (\bibinfo{year}{1995}), \eprint{arXiv:astro-ph/9504054}.

\bibitem[{\citenamefont{{Lewis} et~al.}(2000)\citenamefont{{Lewis},
  {Challinor}, and {Lasenby}}}]{2000ApJ...538..473L}
\bibinfo{author}{\bibfnamefont{A.}~\bibnamefont{{Lewis}}},
  \bibinfo{author}{\bibfnamefont{A.}~\bibnamefont{{Challinor}}},
  \bibnamefont{and}
  \bibinfo{author}{\bibfnamefont{A.}~\bibnamefont{{Lasenby}}},
  \bibinfo{journal}{\apj} \textbf{\bibinfo{volume}{538}}, \bibinfo{pages}{473}
  (\bibinfo{year}{2000}), \eprint{arXiv:astro-ph/9911177}.

\bibitem[{\citenamefont{{Hu}}(2001{\natexlab{a}})}]{HuAng}
\bibinfo{author}{\bibfnamefont{W.}~\bibnamefont{{Hu}}}, \bibinfo{journal}{\prd}
  \textbf{\bibinfo{volume}{64}}, \bibinfo{pages}{083005}
  (\bibinfo{year}{2001}{\natexlab{a}}), \eprint{arXiv:astro-ph/0105117}.

\bibitem[{\citenamefont{{Hu}}(2000)}]{HuHar}
\bibinfo{author}{\bibfnamefont{W.}~\bibnamefont{{Hu}}}, \bibinfo{journal}{\prd}
  \textbf{\bibinfo{volume}{62}}, \bibinfo{pages}{043007}
  (\bibinfo{year}{2000}), \eprint{arXiv:astro-ph/0001303}.

\bibitem[{\citenamefont{{Kesden} et~al.}(2002)\citenamefont{{Kesden}, {Cooray},
  and {Kamionkowski}}}]{2002PhRvD..66h3007K}
\bibinfo{author}{\bibfnamefont{M.}~\bibnamefont{{Kesden}}},
  \bibinfo{author}{\bibfnamefont{A.}~\bibnamefont{{Cooray}}}, \bibnamefont{and}
  \bibinfo{author}{\bibfnamefont{M.}~\bibnamefont{{Kamionkowski}}},
  \bibinfo{journal}{\prd} \textbf{\bibinfo{volume}{66}},
  \bibinfo{pages}{083007} (\bibinfo{year}{2002}),
  \eprint{arXiv:astro-ph/0208325}.

\bibitem[{\citenamefont{{Hu}}(2001{\natexlab{b}})}]{HuMap}
\bibinfo{author}{\bibfnamefont{W.}~\bibnamefont{{Hu}}},
  \bibinfo{journal}{\apjl} \textbf{\bibinfo{volume}{557}}, \bibinfo{pages}{L79}
  (\bibinfo{year}{2001}{\natexlab{b}}), \eprint{arXiv:astro-ph/0105424}.

\bibitem[{\citenamefont{{Perotto} et~al.}(2006)\citenamefont{{Perotto},
  {Lesgourgues}, {Hannestad}, {Tu}, and {Y Y Wong}}}]{Perotto}
\bibinfo{author}{\bibfnamefont{L.}~\bibnamefont{{Perotto}}},
  \bibinfo{author}{\bibfnamefont{J.}~\bibnamefont{{Lesgourgues}}},
  \bibinfo{author}{\bibfnamefont{S.}~\bibnamefont{{Hannestad}}},
  \bibinfo{author}{\bibfnamefont{H.}~\bibnamefont{{Tu}}}, \bibnamefont{and}
  \bibinfo{author}{\bibfnamefont{Y.}~\bibnamefont{{Y Y Wong}}},
  \bibinfo{journal}{Journal of Cosmology and Astro-Particle Physics}
  \textbf{\bibinfo{volume}{10}}, \bibinfo{pages}{13} (\bibinfo{year}{2006}),
  \eprint{arXiv:astro-ph/0606227}.

\bibitem[{\citenamefont{Lewis et~al.}(2011)\citenamefont{Lewis, Challinor, and
  Hanson}}]{Lewis:2011fk}
\bibinfo{author}{\bibfnamefont{A.}~\bibnamefont{Lewis}},
  \bibinfo{author}{\bibfnamefont{A.}~\bibnamefont{Challinor}},
  \bibnamefont{and} \bibinfo{author}{\bibfnamefont{D.}~\bibnamefont{Hanson}}
  (\bibinfo{year}{2011}), \eprint{1101.2234}.

\bibitem[{\citenamefont{{Dvorkin} and {Smith}}(2009)}]{DvoSmi}
\bibinfo{author}{\bibfnamefont{C.}~\bibnamefont{{Dvorkin}}} \bibnamefont{and}
  \bibinfo{author}{\bibfnamefont{K.~M.} \bibnamefont{{Smith}}},
  \bibinfo{journal}{\prd} \textbf{\bibinfo{volume}{79}},
  \bibinfo{pages}{043003} (\bibinfo{year}{2009}), \eprint{0812.1566}.

\bibitem[{\citenamefont{{G{\'o}rski} et~al.}(2005)\citenamefont{{G{\'o}rski},
  {Hivon}, {Banday}, {Wandelt}, {Hansen}, {Reinecke}, and
  {Bartelmann}}}]{GorskHeal}
\bibinfo{author}{\bibfnamefont{K.~M.} \bibnamefont{{G{\'o}rski}}},
  \bibinfo{author}{\bibfnamefont{E.}~\bibnamefont{{Hivon}}},
  \bibinfo{author}{\bibfnamefont{A.~J.} \bibnamefont{{Banday}}},
  \bibinfo{author}{\bibfnamefont{B.~D.} \bibnamefont{{Wandelt}}},
  \bibinfo{author}{\bibfnamefont{F.~K.} \bibnamefont{{Hansen}}},
  \bibinfo{author}{\bibfnamefont{M.}~\bibnamefont{{Reinecke}}},
  \bibnamefont{and}
  \bibinfo{author}{\bibfnamefont{M.}~\bibnamefont{{Bartelmann}}},
  \bibinfo{journal}{\apj} \textbf{\bibinfo{volume}{622}}, \bibinfo{pages}{759}
  (\bibinfo{year}{2005}), \eprint{arXiv:astro-ph/0409513}.

\bibitem[{\citenamefont{{Lewis} et~al.}(2002)\citenamefont{{Lewis},
  {Challinor}, and {Turok}}}]{LewChaTur}
\bibinfo{author}{\bibfnamefont{A.}~\bibnamefont{{Lewis}}},
  \bibinfo{author}{\bibfnamefont{A.}~\bibnamefont{{Challinor}}},
  \bibnamefont{and} \bibinfo{author}{\bibfnamefont{N.}~\bibnamefont{{Turok}}},
  \bibinfo{journal}{\prd} \textbf{\bibinfo{volume}{65}},
  \bibinfo{pages}{023505} (\bibinfo{year}{2002}),
  \eprint{arXiv:astro-ph/0106536}.

\bibitem[{\citenamefont{Newman and Penrose}(1966)}]{NewPenNotes}
\bibinfo{author}{\bibfnamefont{E.~T.} \bibnamefont{Newman}} \bibnamefont{and}
  \bibinfo{author}{\bibfnamefont{R.}~\bibnamefont{Penrose}},
  \bibinfo{journal}{J. Math. Phys.} \textbf{\bibinfo{volume}{7}},
  \bibinfo{pages}{863} (\bibinfo{year}{1966}).

\bibitem[{\citenamefont{{Goldberg} et~al.}(1967)\citenamefont{{Goldberg},
  {Macfarlane}, {Newman}, {Rohrlich}, and {Sudarshan}}}]{GolMacSpinS}
\bibinfo{author}{\bibfnamefont{J.~N.} \bibnamefont{{Goldberg}}},
  \bibinfo{author}{\bibfnamefont{A.~J.} \bibnamefont{{Macfarlane}}},
  \bibinfo{author}{\bibfnamefont{E.~T.} \bibnamefont{{Newman}}},
  \bibinfo{author}{\bibfnamefont{F.}~\bibnamefont{{Rohrlich}}},
  \bibnamefont{and} \bibinfo{author}{\bibfnamefont{E.~C.~G.}
  \bibnamefont{{Sudarshan}}}, \bibinfo{journal}{Journal of Mathematical
  Physics} \textbf{\bibinfo{volume}{8}}, \bibinfo{pages}{2155}
  (\bibinfo{year}{1967}).

\bibitem[{\citenamefont{Varshalovich et~al.}(1988)\citenamefont{Varshalovich,
  Moksalev, and Khersonskii}}]{VarQuant}
\bibinfo{author}{\bibfnamefont{D.~A.} \bibnamefont{Varshalovich}},
  \bibinfo{author}{\bibfnamefont{A.~N.} \bibnamefont{Moksalev}},
  \bibnamefont{and} \bibinfo{author}{\bibfnamefont{V.~K.}
  \bibnamefont{Khersonskii}}, \emph{\bibinfo{title}{Quantum Theory of Angular
  Momentum}} (\bibinfo{publisher}{World Scientific Publishing Co.},
  \bibinfo{year}{1988}).

\bibitem[{\citenamefont{{Hu} and {White}}(1997)}]{HuWhi}
\bibinfo{author}{\bibfnamefont{W.}~\bibnamefont{{Hu}}} \bibnamefont{and}
  \bibinfo{author}{\bibfnamefont{M.}~\bibnamefont{{White}}},
  \bibinfo{journal}{\prd} \textbf{\bibinfo{volume}{56}}, \bibinfo{pages}{596}
  (\bibinfo{year}{1997}), \eprint{arXiv:astro-ph/9702170}.

\end{thebibliography}
